\documentclass[letterpaper,12pt]{JHEP3}
\usepackage{amsmath}
\usepackage{amssymb}
\usepackage{bbm}
\usepackage{subfigure} 
\usepackage{epsfig}
\usepackage{yfonts}
\newcommand{\labell}[1]{\label{#1}}

\newcommand{\be}{\begin{equation}}
\newcommand{\ee}{\end{equation}}
\newcommand{\bea}{\begin{eqnarray}}
\newcommand{\eea}{\end{eqnarray}}
\newcommand{\ba}{\begin{eqnarray}}
\newcommand{\ea}{\end{eqnarray}}

\newcommand{\beq}{\begin{equation}}
\newcommand{\eeq}{\end{equation}}
\newcommand{\beqa}{\begin{eqnarray}}
\newcommand{\eeqa}{\end{eqnarray}}
\newcommand{\beqar}{\begin{eqnarray*}}
\newcommand{\eeqar}{\end{eqnarray*}}

\newcommand{\reef}[1]{(\ref{#1})}

\newcommand{\eg}{{\it e.g.,}\ }
\newcommand{\ie}{{\it i.e.,}\ }

\newcommand{\mt}[1]{\textrm{\tiny #1}}

\newcommand{\E}{\mathcal{E}}
\newcommand{\G}{\mathcal{G}}

\newcommand{\nvec}{{\mbf n}} 

\newcommand{\al}{\alpha}

\newcommand{\lp}{\ell_{\mt P}}
\newcommand{\mbf}{\mathbf}

\newcommand{\qn}{\textswab{q}}
\newcommand{\wn}{\textswab{w}}

\def\tg{{\tilde g}}
\def\ov{\over}
\def\le{\left}
\def\ri{\right}

\preprint{arXiv:1010.0443 [hep-th]}

\title{Holographic Quantum Critical Transport without Self-Duality}

\author{Robert C. Myers,$^{a}$ Subir Sachdev$\,^{b}$ and Ajay Singh$\,^{a,c}$ \\
$^a$ {\it Perimeter Institute for Theoretical Physics, Waterloo,
Ontario N2L 2Y5, Canada}\\
$^b$ {\it Department of Physics, Harvard University, Cambridge MA 02138, USA} \\
$^c$ {\it Department of Physics \& Astronomy and Guelph-Waterloo Physics Institute,}\\
{\it \ \ University of Waterloo, Waterloo, Ontario N2L 3G1, Canada} \\
}
\abstract{We describe general features of frequency-dependent charge
transport near strongly interacting quantum critical points in 2+1 dimensions. The simplest
description using the AdS/CFT correspondence leads to a self-dual
Einstein-Maxwell theory on AdS$_4$, which fixes the conductivity at a frequency-independent
self-dual value. We describe the general structure of higher-derivative corrections to the Einstein-Maxwell theory,
and compute their implications for the frequency dependence of the quantum-critical conductivity.
We show that physical consistency conditions on the higher-derivative terms allow only a limited
frequency dependence in the conductivity.
The frequency dependence is amenable to a physical interpretation using
transport of either particle-like or vortex-like excitations.
}

\begin{document}

\section{Introduction}
\label{sec:intro}

The AdS/CFT correspondence has become a powerful framework for the
study of strongly coupled gauge theories
\cite{Maldacena1,adscft,bigRev}. While it is still in a nascent stage,
an `AdS/Condensed Matter' duality is also being developed. That is, the
AdS/CFT correspondence is proving to be a useful tool to study a range
of physical phenomena which bear strong similarity to those at strongly
coupled critical points in condensed matter systems. A variety of
holographic models displaying interesting properties, including
superfluidity, superconductivity and Hall conductivity, have now been
studied \cite{more}. Further interesting models of various types of
nonrelativistic CFT's have also been constructed \cite{nonrel}.

One advantage of the AdS/CFT correspondence is the `uniformity' of the
holographic approach, \ie a single set of calculations can describe the
system in different disparate regimes (\eg $\omega/T\rightarrow0$
versus $T/\omega\rightarrow0$). This can be contrasted with more
conventional field theory analysis of conformal fixed points
\cite{subir1}. However, a surprising result of the original transport
calculations \cite{sach} was that the frequency dependence was rather
trivial. In particular, the conductivity (at zero momentum) showed no
frequency dependence, \ie it was a constant. The authors of \cite{sach}
traced the origin of this remarkable result to the electromagnetic (EM)
self-duality of the bulk Einstein-Maxwell theory in four dimensions.
Again this holographic result stands in contrast with those from more
conventional field theory analysis \cite{subir1,subir2}.

One perspective on these results is regard them as predictions of the
AdS/CFT analysis on the behavior of
{\em nearly perfect fluids\/}.
Such fluids
are strongly interacting quantum systems, found near scale-invariant
quantum critical points, which respond to local perturbations by
relaxing back to local equilibrium in a time of order $\hbar /(k_B T)$,
which is the shortest possible \cite{subir1}. They are expected to have
a shear viscosity, $\eta$, of order $\eta \sim \hbar s/k_B$ \cite{kss},
where $s$ is the entropy density, and many experimental systems behave
in this manner \cite{physicstoday}. At the same footing, we can then
predict that 2+1 dimensional quantum critical systems with a conserved
charge should have a conductivity which is nearly
frequency-independent. Furthermore, in paired electron systems where
the Cooper pair charge is $2e$, the self-dual value of the conductivity
is \cite{mpaf} $4 e^2/h$, and this is close to the value observed in
numerous experimental systems \cite{kapitulnik}. There has been no
previous rationale why self-duality should be realized in these
experiments, and the AdS/CFT theory of perfect fluids offers a
potential explanation.

Measurements of the frequency dependence of the quantum critical conductivity in two spatial
dimensions have so far been
rather limited \cite{engel,armitage}. Engel {\em et al.} \cite{engel} performed microwave measurements
at the critical point between two quantum Hall plateaus. Their results at the critical point do not
show appreciable $\omega$ dependence as $\hbar \omega $ is scanned through $k_B T$.
However, they did not pay particular attention to the value of the quantum critical conductivity
(they focused mainly on the width of the conductivity peak between the plateaus), and it would be useful
to revisit this more carefully in future measurements. In any case, if confirmed, the AdS/CFT
perspective appears to be the natural explanation for this
weak frequency dependence. Graphene also has characteristics of a quantum-critical system
with moderately strong interactions \cite{graphene}, and its conductivity has been measured \cite{basov,heinz} in
the optical regime where $ \omega \gg T$; a frequency-independent conductivity was found, equal to that
of free Dirac fermions. This is as expected, because the Coulomb interactions are marginally irrelevant
in graphene \cite{graphene}. However, for $\omega \sim T$, the interactions are expected to be more
important, and graphene may well behave like a nearly perfect fluid \cite{graphenevis}. A test of this
hypothesis would be provided by
measurements of the conductivity of graphene in this frequency regime, under conditions in which the
electron-electron scattering dominates over disorder-induced scattering.
There have also been discussions of duality in non-linear transport near quantum critical
points \cite{kivelson,shahar1,shahar2}. Again, there is no natural basis for this in the microscopic theory,
while it can emerge easily from an AdS/CFT analysis \cite{cliff,sondhi}.

Given these motivations, it is clearly useful to understand the robustness
of the AdS/CFT self-duality beyond the classical Einstein-Maxwell theory on AdS$_4$.
As was pointed out in \cite{sach}, in many constructions emerging from
string theory, the Maxwell field would have an effective coupling
depending on a scalar field and the EM self-duality would be lost if
the scalar had a nontrivial profile. From the perspective of the
holographic
CFT, one would be extending the theory by introducing a new scalar
operator, and couplings between the new operator and the original
currents
holographically
dual to the Maxwell field. Further, the nontrivial scalar
profile would indicate that one is now studying physics away from the
critical point as (the expectation value of) the scalar operator will
introduce a definite scale into the problem.

However, we wish to understand the limitations of self-duality, while
remaining at the critical point. For this, a possible approach is to
simply modify the CFT through introducing new higher derivative
interactions in the bulk action for the metric and gauge field, \eg see
\cite{Hofman1,qthydro}. The latter are readily seen to change the
$n$-point functions of current and the stress tensor in the CFT. While
conformal symmetry imposes rigid constraints on the two- and
three-point functions of these operators, they are only determined up
to a finite number of constant parameters, \eg the central charges,
which characterize the particular fixed point theory \cite{ozzy}. These
parameters are reflected in the appearance of dimensionless couplings
in the bulk gravitational theory. Hence, to explore the full parameter
space of the holographic CFT's, one must go beyond studying the
Einstein-Maxwell theory and begin to investigate the effect of higher
derivative interactions in the bulk action. This is the approach which
we examine in the present paper. In particular, we investigate the
effects on the charge transport properties of the
holographic
CFT resulting
from adding a particular bulk interaction coupling the gauge field to
the spacetime curvature -- see eq.~\reef{eqn4}.

Our main results for the frequency dependence of the conductivity
without self-duality are given in Fig.~\ref{fig1}.
\FIGURE{
\includegraphics[width=11cm,height=7cm]{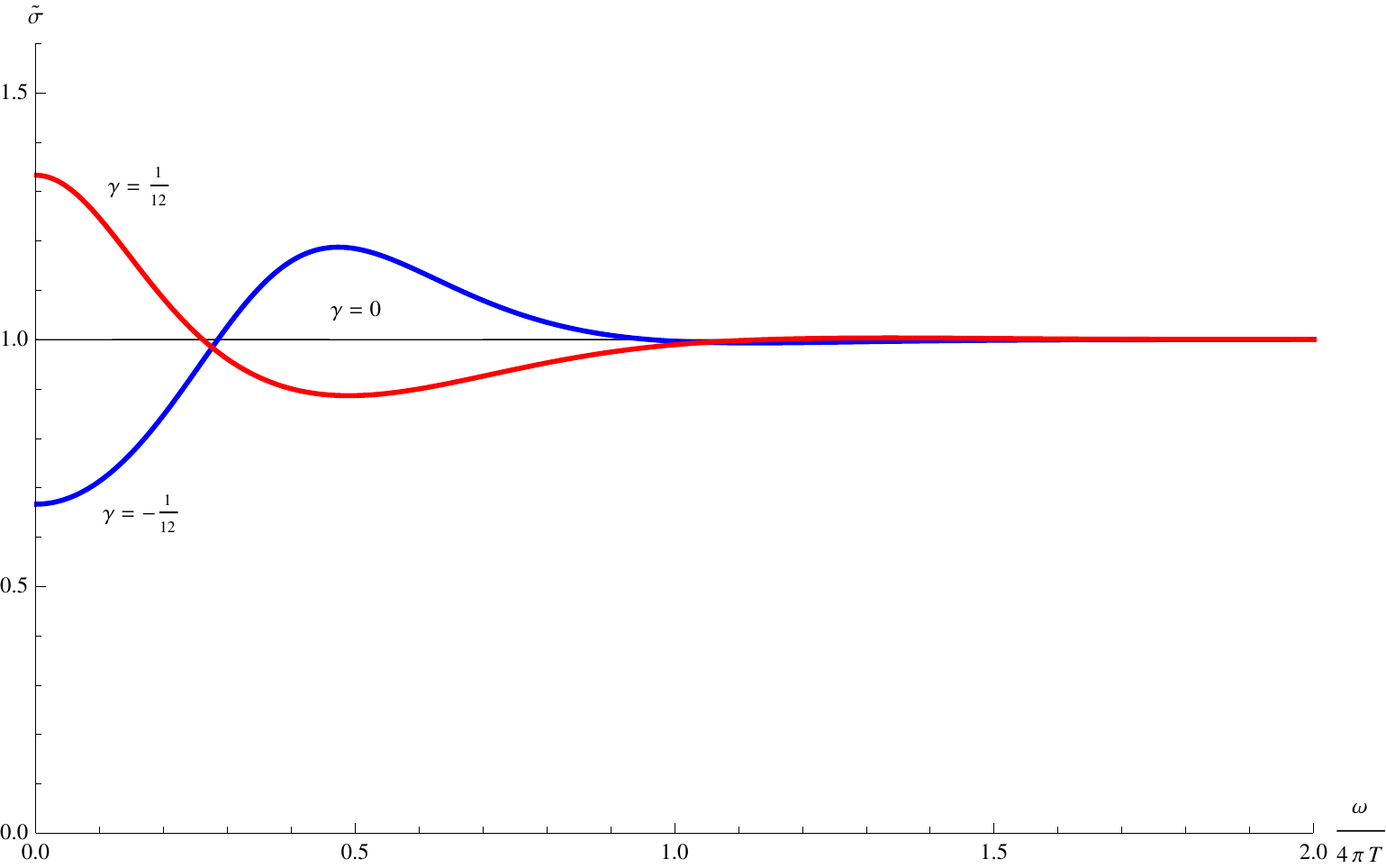}
\caption{The (dimensionless) conductivity $\tilde\sigma=g_4^2\sigma$ is
plotted versus the (dimensionless) frequency $\wn=\omega/(4\pi T)$ for
various values of $\gamma$ (the coupling $g_4$ is defined in
Section~\ref{pree}). Various consistency conditions imply that $\gamma
\in [-1/12,1/12]$ -- see discussion surrounding eq.~\reef{eqn34}. }
\label{fig1}}
Here $\gamma$ is the sole parameter controlling the pertinent higher derivative
terms in the bulk action; we will argue that physical consistency conditions imply
the constraint $|\gamma| < 1/12$.

For $\gamma > 0$, the frequency dependence has the same non-monotonic
form as that expected by extrapolation from the weak-coupling Boltzmann
analysis \cite{subir1}: a collision-dominated Drude peak at small
$\omega$, which is then smoothly connected to the collisionless
$\omega$-independent conductivity at large $\omega$. This similarity
implies that a description of transport in terms of collisions of
charged particles is a reasonable starting point for $\gamma > 0$.

On the other hand, for $\gamma < 0$, we observe that it is the inverse
of the conductivity, \ie the resistivity, which has a Drude-like peak
at small $\omega$. Under particle-vortex duality, the resistivity of
the particles maps onto the conductivity of the vortices \cite{mpaf},
as we will review here in Section~\ref{pavod}. Thus, for $\gamma < 0$,
we conclude that a better description of charge transport is provided
by considering the motion and collisions of {\em vortices\/}. In other
words, for $\gamma < 0$, it is the excitations of the dual holographic
CFT, obtained under the EM duality of the bulk theory, which provide a
Boltzmann-like interpretation of the frequency dependence of the
conductivity.

An outline of the rest paper is as follows: In section \ref{pree}, we
review some basic background material, mainly to motivate the
introduction of the higher derivative interaction for the gauge fields.
In section \ref{diff}, we calculate the charge diffusion constant and
susceptibility for the dual CFT. We turn to the conductivity in section
\ref{cond} and in particular, we demonstrate that in the modified
theory, the conductivity is a nontrivial function of $\omega/T$. In
section \ref{bond}, we derive constraints that arise on the coupling to
the new gauge field interaction by imposing certain consistency
conditions in the dual CFT. We examine electromagnetic duality in the
modified gauge theory in section \ref{duality1}. We conclude with a
brief discussion of our results and future directions in section
\ref{discuss}. A discussion of the Green's functions at finite
frequency and finite momentum is presented in appendix \ref{appA}. In
particular, we examine the relationship between the Green's functions
in the two boundary theories related by EM duality in the bulk.

\section{Preliminaries} \label{pree}

As with many of the recent excursions in the AdS/CMT, our starting
point is the standard Einstein-Maxwell theory (with a negative
cosmological constant) in four dimensions. Hence the action may be
written as
\begin{equation}
I_\mt{0}= \int d^4x \sqrt{-g}\left[ \frac{1}{2\lp^2}
\left(R + \frac{6}{L^2} \right) -\frac{1}{4g_4^2}F_{ab}F^{ab}\right]\,.
\labell{act0}
\end{equation}
The four-dimensional AdS vacuum solution of the above theory
corresponds to the vacuum of the dual three-dimensional CFT. Of course,
the theory also has (neutral) planar AdS black hole solutions:
\begin{equation}
ds^2= \frac{r^2 }{L^2}(- f(r)\,dt^2+dx^2+ dy^2)+\frac{L^2 dr^2}{r^2 f(r)}\,,
\labell{eqn1}
\end{equation}
where $f(r)=1-r_0^3/r^3$. In these coordinates, the asymptotic boundary
is at $r\rightarrow\infty$ and the event horizon, at $r=r_0$. This
solution is dual to the boundary CFT at temperature $T$, where the
temperature is given by the Hawking temperature of the black hole
\begin{equation}
T=\frac{3r_0}{4\pi L^2}\,.
\labell{eqn2}
\end{equation}
At a certain point in the following analysis, it will also be
convenient to work with a new radial coordinate: $u=r_0/r$. In this
coordinate system, the black hole metric becomes
\begin{equation}
ds^2= \frac{r_0^2 }{L^2 u^2}(- f(u)\,dt^2 +dx^2+ dy^2)+\frac{L^2 du^2}{u^2 f(u)}
\,,
\labell{eqn3}
\end{equation}
where $f(u)=1-u^3$. Now the asymptotic boundary is at $u=0$ and horizon
at $u=1$.

As discussed in the introduction, we wish to extend the bulk theory by
adding higher derivative interactions. As usual in quantum field
theory, it is natural to organize the interactions by their dimension
or alternatively by the number of derivatives. The Einstein-Maxwell
action \reef{act0} contains all covariant terms up to two derivatives,
which preserve parity, \ie which are constructed without using the
totally antisymmetric $\varepsilon$ tensor. Hence it is natural to next
consider the possible interactions at fourth order in derivatives
\cite{miguel}. In all, one can construct 15 covariant parity-conserving
terms using the metric curvature, the gauge field strength and their
derivatives \cite{miguel}. However, using integration by
parts,\footnote{Note that we also treat the four-dimensional Euler
density, $R_{abcd}R^{abcd}-4 R_{ab}R^{ab}+R^2$, as trivial since it
does not effect the equations of motion.} as well as the identities
$\nabla_{[a}F_{bc]}= 0= R_{[abc]d}$, the general four-derivative action
can be reduced to eight independent terms
 \beqa
 I_\mt{4}&=&\int d^4x
\sqrt{-g}\left[\,\al_1 R^2+\al_2 R_{ab}R^{ab}+ \al_3
\left(F^2\right)^2+ \al_4 F^4 +\ \al_5 \nabla^aF_{ab}
\nabla^cF_c{}^b\right.
 \nonumber\\
&&\left.\qquad\qquad\qquad\qquad+ \al_6 R_{abcd}F^{ab}F^{cd} + \al_7
R^{ab}F_{ac}F_b{}^c + \al_8 R F^2  \right]
  \labell{act14}
 \eeqa
where $F^2=F_{ab}F^{ab}$, $F^4=F^a{}_bF^b{}_cF^c{}_dF^d{}_a$ and the
$\alpha_{\rm i}$ are some unspecified coupling constants.

In a string theory context, one might expect all of these interactions
to emerge in the low-energy effective action as quantum (\ie
string-loop) or $\al'$ corrections to the two-derivative supergravity
action -- see, for example, \cite{tachi}. In such a context, these
terms would be part of a perturbative expansion where the contribution
of the higher order terms is suppressed by powers of, \eg the ratio of
the string scale over the curvature scale. From the perspective of the
dual conformal gauge theory, these contributions would represent
corrections suppressed by inverse powers of the `t Hooft coupling
and/or the number of colours. Within this perturbative framework, one
is also free to use field redefinitions to simplify the general bulk
action \reef{act14}. In the present case, field redefinitions can be
used to set to zero all of the couplings except three, \eg $\al_3$,
$\al_4$ and $\al_6$ \cite{miguel}. Examining the remaining three terms,
the $\al_3$ and $\al_4$ terms involve four powers of the field strength
and so would not modify the conductivity, at least if we study the
latter at zero density. Hence we are left to consider only the $\al_6$
term which couples two powers of the field strength to the spacetime
curvature. The latter will certainly modify the charge transport
properties of the CFT and, as we discuss in detail in section
\ref{duality1}, it also ruins the EM self-duality of the bulk Maxwell
theory.

While these string theory considerations naturally lead us to focus our
attention on a single new four-derivative interaction, they are limited
to the perturbative framework described above. However, we would also
like to extend our analysis to the case where the new interactions are
making finite modifications of the transport properties. In this case,
we should think of the holographic theory as a toy model whose
behaviour might be indicative of that of a complete string theory
model. Recently the utility of this approach has been shown in
holographic investigations with various higher curvature gravity
theories -- see, for example,
\cite{qthydro,Buchel0,shank,Brigante,highc}. Further, while the
couplings of the higher derivative interactions are finite in this
approach, consistency of the dual CFT prevents these couplings of from
becoming very large, at least in simple models, as we discuss in
section \ref{bond}.

So given this perspective of constructing a toy model with finite
couplings, let us re-examine each of the terms in the general action
\reef{act14}. The first two terms are curvature-squared interactions
which do not involve the gauge field. Hence from the CFT perspective,
these terms would only modify the $n$-point functions of the stress
tensor and so are not relevant to the charge transport. Again, the
third and fourth terms involve four field strengths and so these would
only modify the four-point correlator of the dual current. Hence, as
noted above, these terms will again be irrelevant to the charge
transport, if we limit ourselves to the case of a vanishing chemical
potential. Considering next the $\al_5$ term, we note that it contains
two powers of the field strength and so will modify the charge
transport. However, this term produces higher derivative equations of
motion for the gauge field and so, as explained in detail in
\cite{see}, the dual CFT will contain nonunitary operators. Hence we
discard this term in the analysis at finite coupling to avoid this
problem. Finally, the last two terms in the action \reef{act14} also
involve $F^2$ and again modify the charge transport. However, as we
discuss in more detail in section \ref{discuss}, they only do so in a
trivial way by renormalizing the overall coefficient of the Maxwell
term. Therefore we are again naturally led to consider the $\al_6$
interaction alone in studying the transport properties of dual CFT.

Hence we will study the holographic transport properties with the
following effective action for bulk Maxwell field:
\begin{equation}
I_{vec}=\frac{1}{g^2_4} \int d^4x \sqrt{-g}\left[-\frac{1}{4}F_{ab}F^{ab} +
 \gamma\, L^2  C_{abcd} F^{ab}F^{cd} \right]\,,
\labell{eqn4}
\end{equation}
where we have formulated the extra four-derivative interaction in terms
of the Weyl tensor $C_{abcd}$. That is, it is constructed as a
particular linear combination of the $\al_{6,7,8}$ terms in the general
action \reef{act14}. This particular interaction has the advantage that
it leaves the charge transport at zero temperature unchanged since the
Weyl curvature vanishes in the AdS geometry. Further the factor of
$L^2$ was introduced above so that the coupling $\gamma$ is
dimensionless. From this action, we find the generalized vector
equations of motion:
\begin{equation}
\nabla_{a}\left[F^{ab}- 4\gamma L^2 C^{abcd} F_{cd}\right] = 0\,.
\labell{eqn5}
\end{equation}
Note that the AdS vacuum and (neutral) planar black hole solution
\reef{eqn1} are still solutions of the modified metric equations
produced by the new action.

In closing this discussion, we must note that the four-derivative
interaction in eq.~\reef{eqn4} has also appeared in previous
holographic studies \cite{Hofman1,Hofman2,ritz}. In particular,
\cite{Hofman1,Hofman2} considered the restrictions that must be imposed
on the coupling $\gamma$ in order that the dual CFT is physically
consistent. While \cite{ritz} focused primarily on a five-dimensional
bulk theory, there is considerable overlap between the latter and the
present paper. In particular, \cite{ritz} considered the charge
diffusion constant and (zero-frequency) conductivity, as in section
\ref{diff}, and bounds arising from requiring micro-causality of the
dual CFT, as in section \ref{bond}.

\section{Diffusion Constant and Susceptibility} \label{diff}

In this section, we calculate the charge diffusion constant and
susceptibility, two quantities which control the two point
Green's function of the dual current in the limit of low frequency and
long wavelength \cite{sach}. We follow \cite{shank,ritz} to extend the
analysis of \cite{Kovtun} to accommodate our modified Maxwell action
\reef{eqn4}. We begin by writing a generalized action which is
quadratic in the field strength:
 \be
I= \int d^4 x \, \sqrt{- g}\, \le(-{1 \ov 8g_4^2}\,
F_{ab}\,X^{abcd}\,F_{cd} \ri) \ ,
 \labell{general}
 \ee
where the background tensor $X^{abcd}$ necessarily has the following
symmetries,
 \be
X^{abcd}=X^{[ab][cd]}=X^{cdab}\ .
 \label{symm}
 \ee
The standard Maxwell theory would be recovered by setting
\begin{equation}
X_{ab}{}^{cd}=I_{ab}{}^{cd}=\delta_a{}^c\delta_b{}^d
-\delta_a{}^d\delta_b{}^c\,,
\labell{eqnx2}
\end{equation}
where we can think of $I$ as the identity matrix acting in the space of
two-forms (or anti-symmetric matrices). That is, given an arbitrary
two-form $f_{ab}=-f_{ba}$, then $f_{ab}=\frac12 I_{ab}{}^{cd}f_{cd}$.
With the generalized action in eq.~\eqref{general}, the theory of
interest \eqref{eqn4} is constructed by setting
\begin{equation}
X_{ab}{}^{cd}=I_{ab}{}^{cd} -8\gamma L^2 C_{ab}{}^{cd}\,.
\labell{eqn77}
\end{equation}

Extending the discussion of the membrane paradigm in \cite{Kovtun} to
this generalized framework is straightforward \cite{shank}. One defines
the stretched horizon at $r=r_H$ (with $r_H>r_0$ and $r_H-r_0\ll r_0$)
and the natural conserved current to consider is then
 \be
j^a =  {1\ov4} \le. \,n_b\,X^{abcd}\,F_{cd}\ri|_{r=r_H}\ ,
\label{current}
 \ee
where $n_a$ is an outward-pointing radial unit vector. Then following
the analysis in \cite{Kovtun}, one arrives at the following expression
for the charge diffusion constant \cite{shank}:\footnote{As noted in
\cite{shank}, there are two conditions required for the following
general formulae to hold. The tensor $X_{ab}{}^{cd}$ is: {\it i)}
nonsingular on the horizon and {\it ii)} `diagonal' in the sense
discussed in section \ref{duality1}. Of course, in the present case,
both of these requirements are satisfied by eq.~\reef{eqn77}.
\labell{footy}}
\begin{equation}
D =\le.-\sqrt{-g} \sqrt{-X^{xtxt}\,X^{xrxr}}\ri|_{r=r_0}\
\int_{r_0}^\infty {dr\ov \sqrt{-g}\,X^{trtr}}\ .
 \labell{diffusion}
\end{equation}
Further applying Ohm's law on stretched horizon, the conductivity at
zero frequency is given by \cite{ritz}
 \be
\sigma_0\equiv\sigma(\omega=0,k=0) = \frac{1}{g_4^2}\sqrt{-g}
\sqrt{-X^{xtxt}X^{xrxr}}\big|_{r=r_0}\,. \labell{eqn5x2}
 \ee
Lastly, the susceptibility is easily determined using the Einstein
relation $D=\sigma_0/\chi$. Combining this relation with
eqs.~\reef{diffusion} and \reef{eqn5x2}, an expression for $\chi$ is
easily read off as \cite{ritz}
 \be
\chi^{-1}=-g_4^2\, \int_{r_0}^{\infty}\frac{dr}{\sqrt{-g}X^{trtr}}\,.
 \labell{suss}
 \ee
Of course, if one replaces $X_{ab}{}^{cd}=I_{ab}{}^{cd}$ as in
eq.~\reef{eqnx2}, then these expressions reduce to the expected results
for Einstein-Maxwell theory, \eg see \cite{sach}.

In the present case, we are interested in $X$ as given in
eq.~\reef{eqn77} where the Weyl tensor is evaluated for the planar AdS
black hole \reef{eqn1}. Hence we find
\begin{align}
\sqrt{-g}\sqrt{-X^{xtxt}X^{xrxr}}\big|_{r=r_0} \;=\; &
1+4\gamma \quad \rm{and} \notag \\
\frac{1}{{\sqrt{-g}X^{trtr}}}\;=\;& -\frac{L^2\, r}{r^3-8 r_0^3 \gamma }\,.
\labell{eqn5x5}
\end{align}
Combining these expressions in eq.~\reef{diffusion}, we find the
diffusion constant to be
 \be
D=\frac{ 1+4 \gamma }{16 \pi T\, \gamma^{1/3}}\left(\sqrt{3}\pi-2
\sqrt{3} \arctan\left[ \frac{1+\gamma^{1/3}}{\sqrt{3} \gamma^{1/3}}
\right] + \log \left[ \frac{1- 8\gamma}{ (1-2\gamma^{1/3})^3} \right]
\right)\,.
 \labell{eqn5x7}
 \ee
A plot of this result is given in Fig.~\ref{fig0}. If we consider
$\gamma\ll 1$, this expression simplifies to
 \be
 D \simeq \frac{3}{4\pi T}\left(
1+ 6\gamma + \frac{120}{7}\gamma^2+O(\gamma^3)\,\right)\,.
 \labell{eqn5x8}
 \ee
A perturbative result for $D$ to linear order in $\gamma$ was presented
in \cite{ritz} for arbitrary dimensions and our results above match
that for the case of a three-dimensional CFT.
\FIGURE{
\includegraphics[width=11cm,height=7cm]{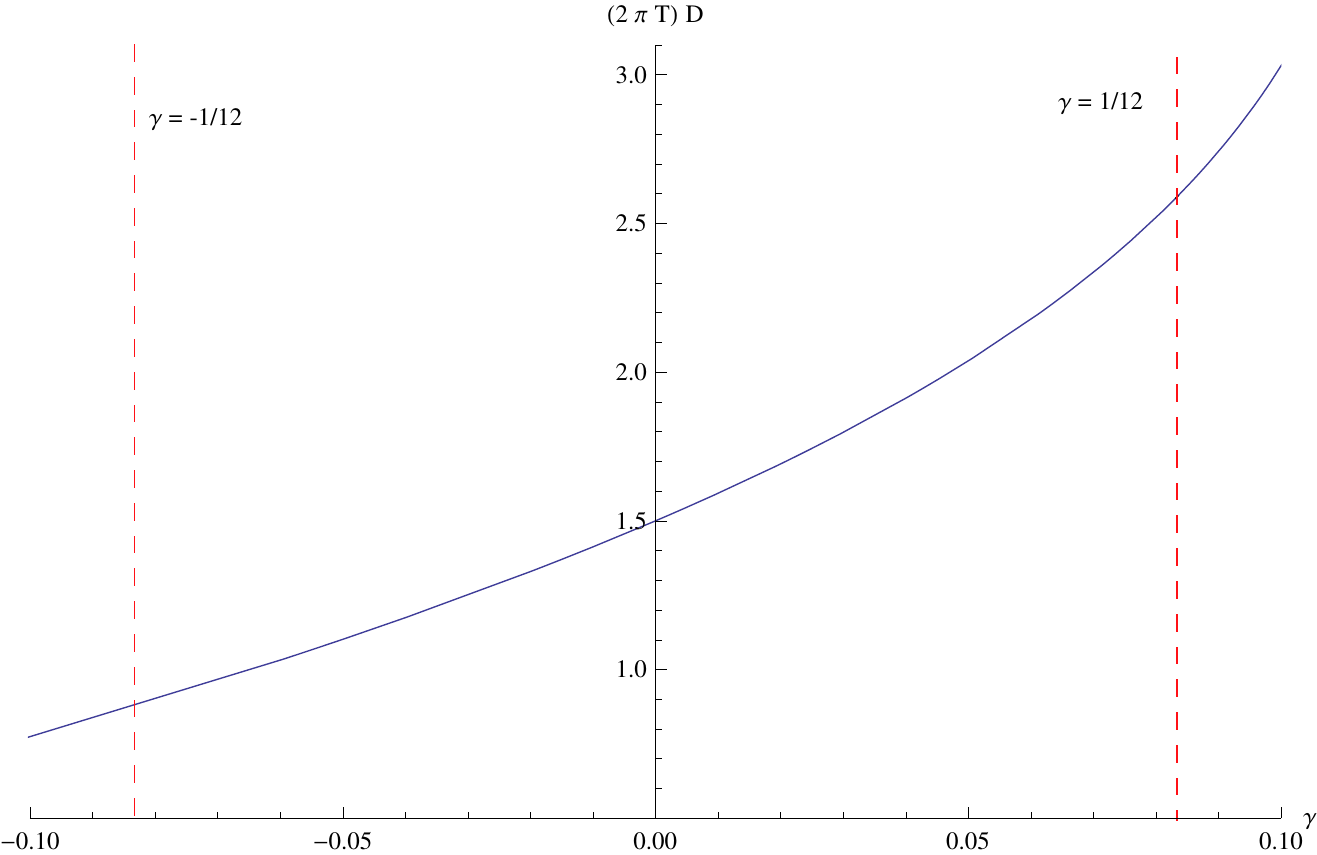}
\caption{The charge diffusion constant is plotted versus the coupling
$\gamma$. The vertical dashed lines denote the boundaries of the
physical regime, $\gamma \in [-1/12,1/12]$ -- see discussion
surrounding eq.~\ref{eqn34}. } \label{fig0}}

Next using \eqref{eqn5x2}, we find
 \be
\sigma_0=\frac{1}{g_4^2}(1+4\gamma)\,. \labell{eqn5x3}
 \ee
Note that this expression is the exact result for arbitrary $\gamma$.
The simple $\gamma$-dependence appearing in the conductivity contrasts
with the complicated formula for the diffusion constant \reef{eqn5x7}.
Of course, the diffusion constant still varies very smoothly with
$\gamma$ in the physical regime, as shown in Fig.~\ref{fig0}. We will
confirm the above result by directly evaluating the two-point function
of the dual current in the next section.

Given these results and the Einstein relation $D=\sigma_0/\chi$, the
susceptibility is easily determined to be
 \be
\chi^{-1}=\frac{ g_4^2}{16 \pi T \,\gamma^{1/3}}\left(\sqrt{3}\pi-2
\sqrt{3} \arctan\left[ \frac{1+\gamma^{1/3}}{\sqrt{3} \gamma^{1/3}}
\right] + \log \left[ \frac{1- 8\gamma}{ (1-2\gamma^{1/3})^3} \right]
\right)\,.
 \labell{chi}
 \ee
Again considering small $\gamma$, the susceptibility reduces to
 \be
  \chi \simeq \frac{4\pi T}{3g_4^2}\left(
1- 2\gamma -\frac{36}{7}\gamma^2+ O(\gamma^3)\,\right)\,.
 \labell{eqn5x9}
 \ee

\section{Conductivity} \label{cond}

In this section, we calculate the conductivity for the CFT dual to the
bulk action \reef{eqn4}. We begin by decomposing the gauge field as
\begin{equation}
A_a(t,x,y,u)=\int \frac{d^3q}{(2\pi)^3} e^{i \text{\bf{q}}\cdot\text{\bf{x}}}
A_a(u,\text{\bf{q}})\,,
\labell{eqn6}
\end{equation}
where $\text{\bf{q}}\cdot\text{\bf{x}}=-\omega t+q^x x+q^y y$. For
convenience and without loss of generality, we choose three-momentum
vector to be $\text{\bf{q}}^{\mu}=(\omega, q, 0)$. Further we choose
the gauge in which $A_u(u,\text{\bf{q}})=0$. Then evaluating modified
Maxwell's equations \reef{eqn5} in the planar black hole background
\reef{eqn3}, we find
\begin{eqnarray}
A_t' + \frac{q f (3-16 u^2 \gamma  f'')}{\omega  (3+32 u^2 \gamma  f'')}
A_x' &= &0 \labell{eqn7}\\
A_t'' + \frac{4 u \gamma  \left(2 f''+u f'''\right)}{3+4 u^2 \gamma  f''} A_t'
- \frac{L^4}{r_0^2} \frac{q  \left(3 - 2u^2 \gamma
f''\right)}{ f \left(3+4 u^2 \gamma  f''\right)} (q A_t+\omega A_x) &=&0 \labell{eqn8}\\
A_x''+\frac{f' (3-2u^2 \gamma  f'')-2u \gamma  f (2 f''+u f''')}{f (3-2u^2 \gamma  f'')}A_x'+
\frac{L^4}{r_0^2} \frac{\omega }{ f^2}
(q A_t+\omega A_x)& =&0 \labell{eqn8x1} \\
A_y''+\frac{f' (3- 2u^2 \gamma  f'')- 2u \gamma  f (2 f''+u f''')}{f
(3- 2u^2 \gamma  f'')} A_y'
\qquad\qquad\qquad\qquad\quad\  && \notag \\
+\frac{L^4}{r_0^2}\frac{ \left(3 \omega ^2-3 q^2 f- 2u^2 \gamma  \left(\omega ^2+2 q^2 f\right)
f''\right)}{f^2 \left(3- 2u^2 \gamma  f''\right)} A_y &=&0 \,. \labell{eqn9}
\end{eqnarray}
Now we can use equations \eqref{eqn7} and \eqref{eqn8} to decouple
equation of motion for $A_t(u,\text{\bf{q}})$:
\begin{equation}
A_t''' + g_1(u) A_t'' + g_2(u) A_t'=0\,,
\labell{eqn10}
\end{equation}
where
\begin{align}
g_1(u)\quad &= \quad \frac{f' (9+6 u^2 \gamma  f''- 64 u^4 \gamma ^2 f''^2 )
+ 2u \gamma  f (15-4 u^2 \gamma  f'') (2 f''+u f''')}{f (3- 2u^2
\gamma  f'') (3+ 2u^2 \gamma  f'')}\,, \notag \\
g_2(u)\quad &= \quad\frac{1}{r_0^2 f^2 (3- 2u^2 \gamma  f'') (3+4 u^2 \gamma  f'')}
\left( L^4 \omega ^2 (9+6 u^2 \gamma  f''-64 u^4 \gamma ^2 f''^2) \right.\notag \\
& \qquad  + f (3- 2u^2 \gamma  f'') (-3 q^2 L^4+ 2u \gamma (q^2 L^4 u+
4 r_0^2 f') f''+4 r_0^2 u^2 \gamma  f' f''') \notag \\
&\left. \qquad  +8 r_0^2 \gamma  f^2 (3 f''+2 u^2 \gamma f''^2+
6 uf'''+ u^4 \gamma  f'''^2 )\right)\,.
\labell{eqn11}
\end{align}

At this point, recall that in the analysis of the Maxwell theory in
\cite{sach}, the equations of motion for $A_y(u,\text{\bf{q}})$ and
$A_t'(u,\text{\bf{q}})$, \ie the $\gamma=0$ limit of eqs.~\eqref{eqn9}
and \eqref{eqn10}, were identical. This was a result of the EM
self-duality of this bulk theory. However, clearly eqs.~\eqref{eqn9}
and \eqref{eqn10} are no longer identical with nonvanishing $\gamma$,
indicating that the new interaction in eq.~\reef{eqn4} breaks the EM
self-duality in the present case. We return to examine the EM duality
in detail in section \ref{duality1}.

Next we solve eq.~\eqref{eqn9} with an infalling boundary condition at
the horizon. Near the horizon, we can write $A_y(u,\text{\bf{q}})=
(1-u)^b F(u,\text{\bf{q}})$ where $F(u,\text{\bf{q}})$ is regular at
$u=1$. Inserting this ansatz in eq.~\eqref{eqn9}, we find that $b=\pm i
{ L^2 \omega }/(3 r_0)$. The ingoing boundary condition at the horizon
fixes
\begin{equation}
b=-i\frac{ L^2 \omega }{3 r_0} = -i\,\wn\,,
\labell{eqn12}
\end{equation}
where we have defined the dimensionless frequency
\begin{equation}
\wn \equiv \frac{\omega}{4 \pi T}\,.
\labell{eqn13}
\end{equation}
As we wish to calculate the conductivity with $\omega\ne0$ but $q=0$
(recall that $q$ is spatial momentum along $x$-direction), we simplify
the notation by denoting $A_a(u,\omega,q=0)$ and $F(u,\omega,q=0)$ by
$A_a(u)$ and $F(u)$. With $b$ given by \eqref{eqn12}, for $q=0$, the
equation of motion for $F(u)$ reduces to
\begin{align}
&0= F''+\left( \frac{3 u^2 (1-4 (1-2 u^3) \gamma )}{(1-u^3) (1+4 u^3 \gamma )}-
\frac{2 i\, \wn }{1-u} \right) F'
\labell{eqn14}\\
&\quad + \frac{i \wn\left(  (1+u+u^2) (1+2 u +2u^2(3+4u+5 u^2) \gamma )-i (2+u)
 (4+u+u^2) (1+4 u^3 \gamma) \wn \right)}{(1-u) (1+u+u^2)^2 (1+4 u^3 \gamma )}F\,. \notag
\end{align}

To proceed further, we need to recall the relation of the conductivity
to the retarded Green's function $\G_{yy}$ for the dual current $J_y$:
\begin{equation}
\sigma = -\text{Im}\left( \frac{\mathcal{G}_{yy}(\text{\bf{q}})}{\omega} \right),
\labell{eqn16}
\end{equation}
Of course, we wish to calculate $\G_{yy}$ using the AdS/CFT
correspondence, following \cite{Son2}. Briefly, integrating by parts in
the action \reef{eqn4}, the bulk contribution vanishes by the equations
of motion \reef{eqn5} and so the result reduces to a surface term. At
the asymptotic boundary, one has the following contribution for $A_y$
 \beqa
I_{yy}&=&-\frac{1}{2g_4^2}\left.\int d^3x \sqrt{-g} g^{uu} g^{yy}\left(
1 -  8\gamma L^2 C_{uy}{}^{uy} \right)
  A_{y}(u,\text{\bf{x}}) \partial_u A_{y}(u,\text{\bf{x}})\right|_{u\to 0}\nonumber\\
  &=&-\frac{2\pi T}{3 g_4^2}\left.\int d^3x  A_{y}(u,\text{\bf{x}})
  \partial_u A_{y}(u,\text{\bf{x}})\right|_{u\to 0}\,.
\labell{eqn18}
 \eeqa
The simple expression in the second line results from explicitly
evaluating the expression with the black hole metric \reef{eqn3} for
which
\begin{equation}
C_{uy}{}^{uy}=-\frac{u^3 }{2L^2}\,.
\labell{eqn19}
\end{equation}
The Fourier transform of $A_y$ is required to compare the above
expression with the standard AdS/CFT result
\begin{equation}
I_{yy}=\int \frac{ d^3q}{(2\pi)^3}\frac{1}{2}  A_{y}(-\text{\bf{q}})
\,\mathcal{G}_{yy}(\text{\bf{q}})\,A_{y}(\text{\bf{q}})\Big|_{u\to0} \,.
\labell{eqn20x3}
\end{equation}
Hence we can arrive at the usual result, \ie the coupling $\gamma$
makes no explicit appearance here,
\be
\mathcal{G}_{yy}(\text{\bf{q}})=-\frac{4\pi T}{3 g_4^2}\left.
\frac{A_{y}(u,-\text{\bf{q}})\,\partial_u A_{y}(u,\text{\bf{q}})
}{A_{y}(u,-\text{\bf{q}})\, A_{y}(u,\text{\bf{q}}) } \right|_{u\to0}\,.
 \labell{eqn20x4}
\ee
Focusing our attention on the case $\text{\bf{q}}^{\mu}=(\omega,0,0)$
and adopting the notation introduced above eq.~\reef{eqn14}, the
retarded Green's function becomes
\begin{equation}
\mathcal{G}_{yy}(\omega,q=0)=-\frac{4\pi T}{3 g_4^2} \frac{\partial_u
A_{y}(u,\omega)}{A_y(u,\omega)} \Big|_{u\to0}\,.
\labell{eqn21}
\end{equation}
Then eq.~\eqref{eqn16} yields the conductivity at $q=0$ as
\begin{equation}
\sigma=\frac{1}{3 g_4^2} \text{Im}\left( \frac{\partial_u A_{y}}{\wn A_y} \right)_{u\to0}\,.
\labell{eqn22}
\end{equation}

Given the above expression, it is straightforward to calculate
conductivity for small $\omega$ analytically and confirm the result
\reef{eqn5x3} for $\sigma_0=\sigma(\omega=0,q=0)$ derived in the
previous section using the membrane paradigm. First, we make a Taylor
expansion of $F(u)$ in $\wn$ and substitute the ansatz $F(u)=F_1(u) +
\wn F_2(u)$ into \eqref{eqn14}. Then, we find that $F_1$ and $F_2$
should satisfy the following
\begin{align}
F_1''-\frac{3 u^2 (1-4(1-2 u^3) \gamma )}{(1-u^3) (1+4 u^3 \gamma )} F_1'& = 0, \labell{eqn22x1} \\
F_2''-\frac{3 u^2 (1-4(1-2 u^3) \gamma)}{ (1-u^3) (1+4 u^3 \gamma)}F_2'
+\frac{2i}{1-u} F_1' \qquad\qquad & \notag \\
+\,\frac{i (1+2 u+4 u^2 \gamma(3 +4 u +5 u^2) )}{(1-u^3) (1+4 u^3 \gamma )} F_1 & =0\,.
\labell{eqn22x2}
\end{align}
After solving eq.~\eqref{eqn22x1} for $F_1$, we can fix one of the
integration constants demanding that $F_1$ is regular at the horizon.
This yields $F_1(u)= C$, where $C$ is an arbitrary constant. Given
$F_1$, we solve eq.~\eqref{eqn22x2} for $F_2$. In this case, we fix the
two integration constants by imposing the following two conditions:
First, $F_2$ is regular at the horizon. Second, we normalize $F(u)$
such that its value at the horizon is independent of $\wn$, \ie
$F_2(u=1)=0$. The final result is given by
\begin{align}
F_2(u)\;=\; & -i C\Bigg( \frac{\pi}{\sqrt{3}} - \sqrt{3} \arctan\left[\frac{1+2 u}{\sqrt{3}}\right]
-\frac12 \log \left[ \frac{1 + u + u^2}{3}\right]
 \labell{eqn22x3} \\
& +2 \sqrt{3}\, 2^{1/3} \gamma^{2/3}\left( \arctan\left[\frac{1-2\,2^{2/3} u \gamma^{1/3}}{\sqrt{3}}\right]
 -\arctan\left[\frac{1-2 \, 2^{2/3} \gamma^{1/3}}{\sqrt{3}}\right]\right)
\notag \\
& +  2^{1/3} \gamma^{2/3} \log\left[ \frac{(1+2^{2/3}\gamma^{1/3})^3}{1+ 4 \gamma}\right] - 2^{1/3}
  \gamma^{2/3} \log\left[\frac{(1+u \, 2^{2/3} \gamma^{1/3})^3}{1+4 u^3\gamma}\right] \Bigg)\,.
\notag
\end{align}
Now we can simply use $A_y(u)\simeq(1-u)^b (F_1(u)+\wn F_2(u))$ in
eq.~\eqref{eqn22}, take the limit $\wn \rightarrow 0$ and find
\begin{equation}
\sigma_0=\frac{1}{ g_4^2} (1+4\gamma)\,,
\labell{eqn22x4}
\end{equation}
which agrees with our previous result \reef{eqn5x2}.

To study frequency dependant conductivity, we must solve
eq.~\eqref{eqn14} numerically. Our numerical integrations run outward
from the horizon and so we need to fix the initial conditions at $u=1$.
To determine the latter we solve eq.~\eqref{eqn14} for $u\ll 1$,
finding
\begin{equation}
F(u)= 1 - (1-u) \frac{i \wn  ( i+2\wn +8 \gamma  (2i+\wn ))}{
(1+4 \gamma ) ( i+\wn )}\,.
\labell{eqn22x5}
\end{equation}
Numerical integration is used to determine $F(u)$ out to the boundary
at $u=0$ for fixed values of $\wn$ (and $\gamma$) and then we use the
complete solution $A_y(u)= (1-u)^b F(u)$ and eq.~\eqref{eqn22} to
calculate conductivity $\sigma (\wn)$. In figure \ref{fig1}, we show
our results for various values of coupling constant $\gamma$.

\section{Bounds on the Coupling} \label{bond}

In this section, we find the constraints that are imposed on the
coupling $\gamma$ by demanding that the dual CFT respects causality,
following the analysis described in \cite{Brigante,Buchel0,ritz}. We
also examine if there are any unstable modes of the vector field, as
discussed in \cite{Myers0,shank}, which would result in our
calculations of the charge transport properties being unreliable. From
a dual perspective, such unstable modes indicate that the uniform
neutral plasma is an unstable configuration in the dual CFT.

To examine causality, the first step is to re-express the equations of
motion of the two independent vector modes, \ie eqs.~\eqref{eqn9} and
\eqref{eqn10}, in the form of the Schr\"odinger equation. We begin by
considering eq.~\eqref{eqn10}. Recall that we are working in the gauge
where $A_u(u,\text{\bf{q}})=0$ and we have chosen
$\text{\bf{q}}^{\mu}=(\omega,q,0)$. Now if we make a coordinate
transformation to $z(u)$ such that
\begin{equation}
z'=\frac{3}{1-u^3}\,,
\labell{eqn23}
\end{equation}
and write $A_t'(u,\text{\bf{q}})= G_1(u)\,\psi_1(u,\text{\bf{q}})$
where
\begin{equation}
G_1'(u)-\frac{6 u^2 \gamma  \left(5+8 u^3 \gamma \right)}{1-4 u^3
\gamma(1 -64 u^3 \gamma)} G_1(u)=0\,,
\labell{eqn24}
\end{equation}
then eq.~\eqref{eqn10} takes the form
\begin{equation}
-\partial_z^2\psi_1(z) + V(z) \psi_1(z) = \wn^2 \psi(z)\,.
\labell{eqn25}
\end{equation}
In this Schr\"odinger form, the effective potential $V(z)$ can be
expressed in terms of $u$ as
\begin{equation}
V(u)=\qn^2 V_0(u) + V_1(u)\,,
\labell{eqn26}
\end{equation}
where
\begin{align}
\qn\quad&\equiv\quad \frac{q}{4 \pi T}\\
V_0(u)\quad &=\quad \frac{(1-u^3) (1+4 u^3 \gamma)}{(1-8 u^3 \gamma)} \labell{eqn27} \\
V_1(u)\quad &= \quad -\frac{ 2u (1-u^3) \gamma  (2- 2u^6 \gamma -5 u^3
 (1+ 2\gamma ))}{3(1+4 u^3 \gamma)^2}\,.
\labell{eqn28}
\end{align}

It is easiest to consider the limit $\qn\to\infty$, in which case one
can solve for $\psi_1$ in a WKB approximation \cite{Brigante}. In this
limit, $V_0(u)$ will dominate the potential and we want to examine how
its properties change as $\gamma$ is varied, \eg following
\cite{Buchel0}. In Fig.~\ref{fig2}, we have plotted potential $V_0(u)$
for various values of $\gamma$. We observe that if $\gamma$ is too
large, the potential develops a maximum with $V_{0,max}>1$ at some
point between $u=0$ and $u=1$. In that case, there will be
`super-luminal' modes with $\wn/\qn=\omega/q>1$ indicating that
causality is violated in the dual CFT \cite{Brigante, Buchel0}. One can
easily verify that this new maximum appears for $\gamma
> 1/12$ by examining the behaviour of $V_0(u)$ near the boundary, \ie near $u=0$,
where eq.~\reef{eqn27} yields
 \be
V_0(u)\simeq 1-\left(1-12\gamma\right)u^3+\cdots\,. \labell{eqn27a}
 \ee
\FIGURE{
\includegraphics[width=0.48\textwidth]{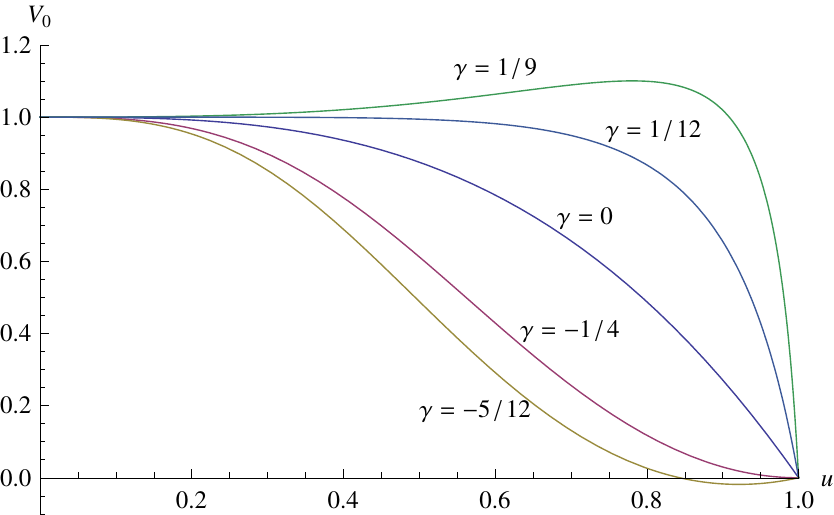}
\includegraphics[width=0.48\textwidth]{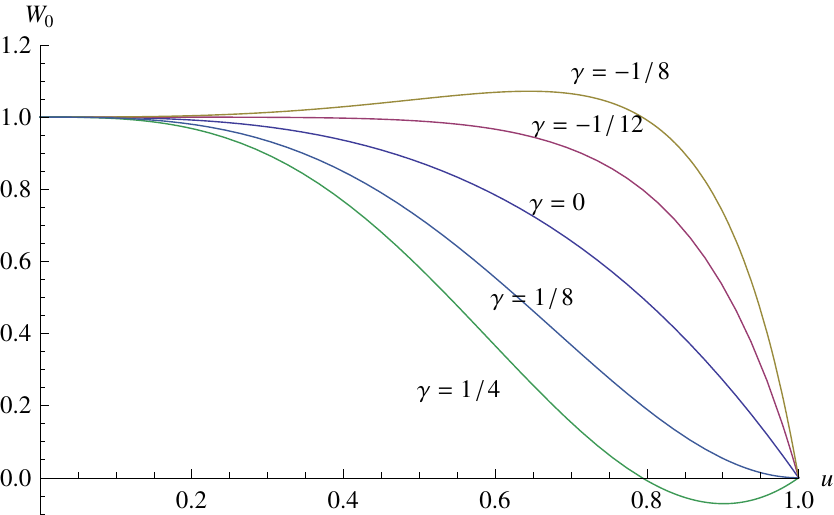}
\caption{Left: $V_0(u)$ for longitudinal $A_t$ mode for various values
of $\gamma$. Right: $W_0(u)$ for transverse $A_y$ mode for various
values of $\gamma$. Consistency conditions discussed in the text for
the longitudinal mode imply $ \gamma \in [-1/4, 1/12]$ in $V_0$.
Similarly for transverse mode, $ \gamma \in [-1/12, 1/8]$ in $W_0$.}
\label{fig2}}

Next we turn to the transverse vector mode satisfying eq.~\eqref{eqn9}.
As above, we make a change of coordinate to $z(u)$ satisfying
eq.~\eqref{eqn23} and we write $A_y(u)= G_2(u)\,\psi_2(u)$ where
\begin{equation}
G_2'(u) + \frac{6 u^2 \gamma }{1+4 u^3 \gamma } G_2(u)=0 \,.
\labell{eqn29}
\end{equation}
With these choices, eq.~\eqref{eqn9} reduces to the desired
Schr\"odinger form
\begin{equation}
-\partial^2_z\psi_2(z) + W(z) \psi_2(z) =  \wn^2 \psi_2(z)\,,
\labell{eqn30}
\end{equation}
where
\begin{align}
W(u) \quad &= \quad \qn^2 W_0(u) + W_1(u) \labell{eqn31} \qquad \text{with}\\
W_0(u)\quad &= \quad \frac{(1-u^3) (1-8 u^3 \gamma )}{(1+4 u^3 \gamma )} \labell{eqn32} \,, \\
W_1(u)\quad &= \quad \frac{ 2u (1-u^3) \gamma (2-5u^3 + 2\gamma u^3(1 -7 u^3) )}{3(1+4 u^3 \gamma)^2}\,.
\labell{eqn33}
\end{align}
We again consider the WKB limit where $W_0$ dominates the potential.
The shape of this potential is also shown in Fig.~\ref{fig2} for
various values of $\gamma$. Examining the potential \reef{eqn32} as
above, we find that a maximum develops for $\gamma < -1/12$, indicating
that causality is violated in the dual CFT in this regime.

Combining the results from both modes, we find that the dual CFT is
only consistent (\ie respects causality) if
\begin{align}
-\frac{1}{12} \leq \gamma \leq \frac{1}{12}\,.
\labell{eqn34}
\end{align}
We also note that these bounds coming from the violation of
micro-causality precisely match the bounds derived for the dual
parameter in the CFT derived in \cite{Hofman1, Hofman2}. There, various
thought experiments were proposed to constrain CFT's in four
dimensions. However, their discussion is readily adapted to the three
dimensions, as we consider here. The relevant experiment consists of
first producing a disturbance, which is localized and injects a fixed
energy, with an insertion of the current $\varepsilon^{i}\,J_i$, where
$\varepsilon^{i}$ is a constant (spatial) polarization tensor. Then one
measures the energy flux escaping to null infinity in the direction
indicated by a unit vector $\nvec$:
 \be
\E(\nvec)=\lim_{r\to +\infty} r \int_{-\infty}^{+\infty} dt~
T^{t}{}_{i}(t,r\, \mbf n )\ n^i\,.\labell{flux0a}
 \ee
The final result takes the form
 \bea
\langle \mathcal{E}(\nvec) \rangle &=& \frac{\langle
0|(\varepsilon^*\cdot j^\dagger)\, \mathcal{E}(\nvec)\, (\varepsilon
\cdot J )|0\rangle}{\langle 0|(\varepsilon^*\cdot J^\dagger) \,
(\varepsilon \cdot j )|0\rangle} =\frac{E}{2\pi} \le[ 1 + a_2 \le(
\frac{|\varepsilon\cdot\nvec|^2}{|\varepsilon|^2} - \frac{1}{2} \ri)
\ri]
 \nonumber\\
&=&\, \frac{E}{2\pi} \le[ 1 + a_2 \le( \cos^2 \theta - \frac{1}{2} \ri)
\ri]\,,
 \labell{eqn34x1}
  \eea
where $E$ is the total energy and $\theta$ is the angle between the
direction $\nvec$ and the polarization $\varepsilon$. The structure of
this expression is completely dictated by the symmetry of the
construction and the (constant) coefficient $a_2$ is a parameter which
characterizes the underlying CFT. Given eq.~\reef{eqn34x1}, it is clear
that $a_2$ is related to the parameters appearing in the general
three-point correlator $\langle T_{ab}(x) J_c(y) J_d(z)\rangle$ -- see
discussion in section \ref{discuss}. Now, the interesting observation
of \cite{Hofman1} was that if the coefficient $a_2$ becomes too large,
the energy flux measured in various directions will become negative.
Hence demanding that the energy flux should be positive in all
directions for a consistent CFT leads to the constraints
 \be
-2 \leq a_2 \leq 2 \,.
 \labell{eqn34x2}
 \ee
Of course, to relate this result to that in eq.~\reef{eqn34}, we must
find the relation between $a_2$ for our holographic CFT and the bulk
coupling $\gamma$. The simplest approach is to use the AdS/CFT
correspondence to examine the bulk dual of the thought experiment
presented above. As noted above, in calculating the flux expectation
value in eq.~\reef{eqn34x1}, we are essentially determining a specific
component of the  three-point function of the stress tensor with two
currents. Hence in our holographic description, we must introduce an
appropriate metric fluctuation $h_{\mu\nu}$ and two gauge field
perturbations $A_\mu$ in the AdS$_4$ bulk, which couple to the boundary
insertions of $T_{ab}$ and $J_a$. We then evaluate the on-shell
contribution for these three insertions with the action \reef{eqn4}. We
do not present the details here, as the analogous calculations for
$d=4$ are presented in Appendix D of \cite{Hofman1} -- the interested
reader may also find the discussion in the first reference in
\cite{qthydro} useful. In the end, the holographic calculations yield a
very simple final result
 \be a_2\,=\,-24 \gamma  \labell{eqn34x4} \ee
and hence we find the bounds in eqs.~\reef{eqn34} and \reef{eqn34x2}
are equivalent.

Next we turn to possible instabilities in the neutral plasma. If we
examine the potential $V_0$ in more detail, we find that another
interesting feature develops for $\gamma < -1/4$. That is, the
potential develops a minimum at some radius close to the horizon where
$V_0(u)<0$. The appearance of this potential well can be verified
analytically by expanding $V_0(u)$ near  $u=1$,
 \be
V_0(u)\simeq 3\frac{1+4\gamma}{1-8\gamma}(1-u)+\cdots\,.
\labell{eqn27b}
 \ee
While $V_0$ always vanishes at $u=1$, we see that for $\gamma < -1/4$,
$V_0<0$ immediately in front of the horizon indicating the presence of
the negative potential well there. In the WKB limit, this potential
well leads to bound states with a negative (effective) energy, which
correspond to unstable quasinormal modes in the bulk theory
\cite{Myers0}. While these modes do not signal a fundamental pathology
with the dual CFT, they do indicate that the uniform neutral plasma is
unstable in this regime. Hence our calculation of the conductivity
would be unreliable here. Of course, our previous constraints
\reef{eqn34} have already ruled out $\gamma < -1/4$ as being physically
interesting and so we need not worry about these instabilities.

On the other hand, one may worry that additional instabilities will
appear outside of the WKB regime, considered above. In particular for
small momentum, the effective potential will also receive an important
contribution from $V_1(u)$. We find that for $ \gamma \in (-1/4,0)$,
$V_1(u)$ also develops a negative minimum close to the horizon and so
there might be some unstable modes in the plasma in this regime as
well. We have plotted the potential $V_1(u)$ for various values of the
coupling constant $\gamma$ in Fig.~\ref{fig2x1}. While the WKB
approximation may be less reliable in this regime, the analysis in
\cite{Myers0} suggests that it is sufficient to determine the
appearance of unstable modes. According to WKB approximation, a zero
energy bound state can appear in this potential well for
\begin{align}
\left( n-\frac{1}{2} \right) \pi \;\simeq\; & \int_{z_0}^{\infty} dz\sqrt{-V_1(z)} \notag \\
= \; & \int_{u=u_0}^{u=1} \frac{3 du}{(1-u^3)} \sqrt{-V_1(u)}\equiv I\,,
\labell{eqn28x1}
\end{align}
where $n$ is a positive integer and the integration is over the values
of $u$ for which the potential is negative. A plot of $\tilde n\equiv
I/\pi+1/2$ is given in Fig.~\ref{fig2x1}. We see that $\tilde n$
reaches a maximum value of approximately 0.86, implying that the
potential well is never able to support a negative energy bound state.
Hence we conclude that there are no unstable modes in this low momentum
regime.
\FIGURE{
\includegraphics[width=0.48\textwidth]{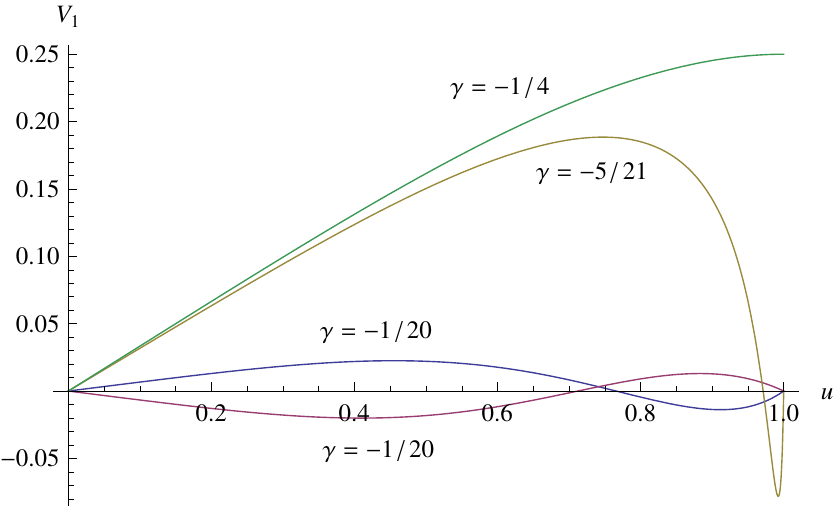}
\includegraphics[width=0.48\textwidth]{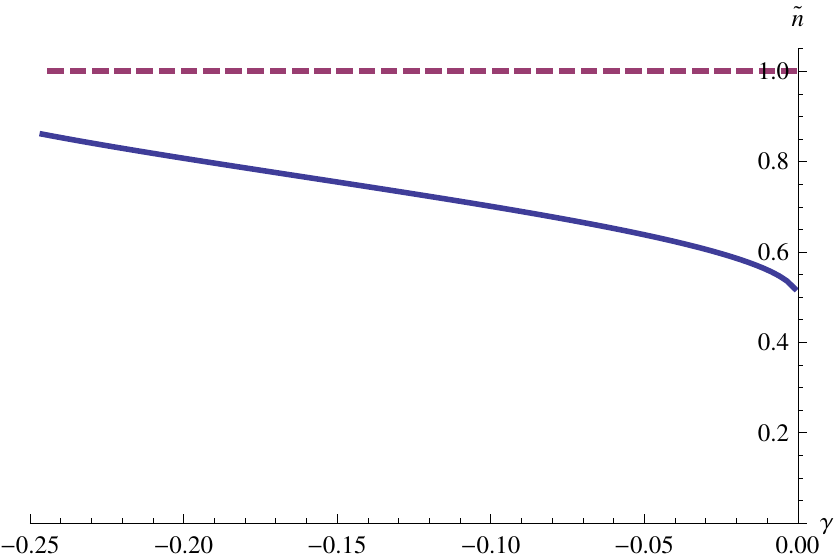}
\caption{Left: $V_1(u)$ for various values of $\gamma$. Right: $\tilde
n=I/\pi+1/2$ plotted versus $\gamma$ -- see eq.\reef{eqn28x1}. Here
potentials $V_1(u)$ is plotted for various values of $\gamma$. We see
that a negative dip appears in $V_1(u)$ close to the horizon for $
\gamma \in (-1/4, 0)$. We have also plotted $\tilde n$ in this range of
$\gamma$ and the plot clearly indicates that it always remains less
than one.} \label{fig2x1}}

While we have discussed both small and large momenta limit of our
effective potential $V(u)$, one may still imagine that instabilities
can still arise at some finite momenta. However, such a possibility can
be eliminated by considering the structure of our complete potential
$V(u)$. That is, for any finite momenta and for
$\gamma \in [-1/4,0]$, the negative dip in potential $V(u)$ is smaller
than the dip in $V_1(u)$ because of the positive contribution coming
from $V_0(u)$. Hence there are no instabilities coming from the
longitudinal vector mode in the regime \reef{eqn34} of physical
interest.

Of course, one must also consider possible instabilities in the
transverse vector mode. In this case, examining the potential $W_0$, we
find that a negative minimum again develops for $\gamma > 1/8$. So
again instabilities appear in the large momentum limit but only for
values of the coupling outside of the physical regime \reef{eqn34}. As
above, one can also consider the low and finite momentum regimes,
however, again one finds that there are no additional instabilities in
the physical regime. Hence although both the transverse and
longitudinal modes of the vector exhibit instabilities, these only
appear in a regime where our previous constraints already indicate that
the CFT is pathological.

Examining eq.~\reef{eqn27b}, one sees that the potential $V_0$ is also
negative in front of the horizon for $\gamma>1/8$ (as well as for
$\gamma<-1/4$, as discussed above). However, this behaviour is not
indicative of a negative potential well in this case. Rather a closer
examination of the full potential \reef{eqn27} shows that a simple pole
appears at $u=1/(2\gamma^{1/3})$, which lies in the physical interval
$0\leq u\leq 1$ for $\gamma>1/8$. The potential $W_0$ exhibits a
similar behaviour for $\gamma<-1/4$. The analysis and physical
interpretation of the modes in this case are more elaborate along the
lines of that given in \cite{qthydro}. However, we do not consider
these issues further here since our previous constraints \reef{eqn34}
already indicate that $\gamma > 1/8$  and $\gamma<-1/4$ are outside of
the physically viable regime.

\section{EM Self-Duality Lost} \label{duality1}

In this section, we examine in more detail the loss of electromagnetic
(EM) self-duality for the U(1) gauge theory defined by the bulk action
(\ref{eqn4}). Recall from \cite{sach} that this EM self-duality was the
key property of the standard four-dimensional Maxwell theory which lead
to the simple relation:
\begin{equation}
K^T(\omega,q)\,K^L(\omega,q)= {\rm constant}\,,
\labell{eqnx1}
\end{equation}
where $K^T$ and $K^L$ are the scalar functions determining the
transverse and longitudinal components of the retarded current-current
correlator -- see appendix \ref{appA} for further discussion. As a
result, the conductivity (at zero momentum) was a fixed constant for
all values of $\omega/T$. In examining the explicit equations of
motion, \reef{eqn9} and \reef{eqn10}, we already noted that
self-duality is lost in the new theory. However, in the context of any
$U(1)$ gauge theory, one can think of EM duality as simply a change of
variables in the corresponding path integral. Even if our new gauge
theory \eqref{eqn4} is not self-dual, we can still implement this
change of variables and construct the EM dual theory, as we will
demonstrate below.

We begin by introducing a (vector) Lagrange multiplier $B_a$ in the
generalized action \reef{general} as follows
\begin{equation}
I= \int d^4x \sqrt{-g}\left(-\frac{1}{8 g^2_4}F_{ab} X^{abcd}F_{cd} +\frac{1}{2}
\varepsilon^{abcd} B_a \partial_b F_{cd} \right)\,.
\labell{eqn67}
\end{equation}
Here $\varepsilon_{abcd}$ is totally antisymmetric tensor, with
$\varepsilon_{0123}=\sqrt{-g}$. The fundamental fields in the path
integral for this action are the two-form $F_{ab}$ and the one-form
$B_a$. Now the EM duality comes from simply treating the integration
over these fields in two different orders.

If we evaluate the path integral by first integrating over the Lagrange
multiplier $B_a$, the latter integration enforces the Bianchi identity
on the two-form $F_{ab}$, \ie
\begin{equation}
\varepsilon^{abcd} \partial_b F_{cd}=0\,.
\labell{eqn70}
\end{equation}
If $F_{ab}$ is to satisfy this constraint,\footnote{Note that we are
justified in using ordinary (rather than covariant) derivatives both
here and in the action \eqref{eqn67} because of the antisymmetry of the
indices.} then on a topologically trivial background, it must take the
form $F_{ab}=\partial_a A_b-\partial_b A_a$. Hence the remaining path
integral reduces to the `standard' gauge theory where the fundamental
field is the Maxwell potential $A_a$ with generalized action given in
eq.~\reef{general}.

Alternatively, one can perform the path integral over the two-form
$F_{ab}$ first. In this case, we first integrate by parts in the second
term in the action \eqref{eqn67}
\begin{equation}
I= \int d^4x \sqrt{-g}\left(-\frac{1}{8 g^2_4}F_{ab} X^{abcd}F_{cd} +\frac{1}{4}
\varepsilon^{abcd} F_{ab}G_{cd} \right)\,.
\labell{eqn67x}
\end{equation}
where we have defined the new field strength
$G_{ab}\equiv\partial_aB_b-\partial_bB_a$. We can now shift the
original two-form field to
\begin{equation}
\widehat F_{ab}=F_{ab}-\frac{g^2_4}{4}\left(X^{-1}\right)_{abcd}
\varepsilon^{cdef}G_{ef}
\labell{shift}
\end{equation}
where $X^{-1}$ is defined by
\begin{equation}
\frac12\,(X^{-1})_{ab}{}^{cd}\,X_{cd}{}^{ef}\equiv I_{ab}{}^{ef}\,.
\labell{inverse}
\end{equation}
Recall the definition of $I_{ab}{}^{cd}$ given in eq.~\reef{eqnx2}.
With this shift, one has a trivial Gaussian integral over the field
$\widehat F_{ab}$ after which one is left with the path integral over
the one-form $B_a$ with the action
\begin{equation}
I= \int d^4x \sqrt{-g}\left(-
\frac{1}{8\hat g_4^2}
\widehat X^{abcd} G_{ab}G_{cd} \right)\,,
\labell{eqn67xx}
\end{equation}
where $\hat g_4^2\equiv 1/g_4^2$ and
\begin{eqnarray}
\widehat X_{ab}{}^{cd}&=&-\frac14\, \varepsilon_{ab}{}^{ef}
\,(X^{-1})_{ef}{}^{\,gh}\,\varepsilon_{gh}{}^{cd} \notag \\
&=&(X^{-1})_{ab}{}^{cd}+\frac12\,
(X^{-1})_{ef}{}^{ef}\,I_{ab}{}^{cd} \labell{kinetic}\\
&&\qquad-\left[(X^{-1})_{ae}{}^{ce}\delta_b{}^d
-(X^{-1})_{ae}{}^{de}\delta_b{}^c
-(X^{-1})_{be}{}^{ce}\delta_a{}^d
+(X^{-1})_{be}{}^{de}\delta_a{}^c
\right]\,.\nonumber
\end{eqnarray}
In the second equality above, the two $\varepsilon$-tensors have been
eliminated with the four-dimensional identity of the form
$\varepsilon_{abcd}\,\varepsilon^{efgh}=-
(\delta_a{}^e\,\delta_b{}^f\,\delta_c{}^g\,\delta_d{}^h+\cdots)$ --
note that the overall minus sign appears because we are working in
Minkowski signature. Hence $B_a$ now plays the role of the gauge
potential in the EM dual theory with the action \eqref{eqn67xx}.

The relation between the gauge fields in the two dual EM theories is
implicit in the equations of motion for $F_{ab}$ or $\widehat F_{ab}$.
From eq.~\eqref{shift}, we see that setting $\widehat F_{ab}=0$ yields
\begin{equation}
F_{ab}=\frac{g^2_4}{4}\left(X^{-1}\right)_{ab}{}^{cd}\,
\varepsilon_{cd}{}^{ef}\,G_{ef}\,.
\labell{duality}
\end{equation}
Recall that in the usual Maxwell theory, $X$ takes the simple form
given in eq.~\eqref{eqnx2}. In this case, $X^{-1}=X$ and one can easily
show that eq.~\eqref{kinetic} also yields $\widehat X_{ab}{}^{cd} =
I_{ab}{}^{cd}$. Hence for the Maxwell theory, the form of the two
actions, \eqref{general} and \eqref{eqn67xx}, as well as the
corresponding equations of motion for $A_a$ and $B_a$, are identical.
This is then a demonstration that the Maxwell theory is self-dual.
Further, the duality relation between the two field strengths in
eq.~\eqref{duality} corresponds to the usual Hodge duality, as expected
for this case.

Of course, in general, we will find that $\widehat X\not = X$ and so
this self-duality property is lost. That is, the form of the action and
the equations of motion in the original theory and its dual now have
different forms, \ie
\begin{equation}
\nabla_b\left(X^{abcd}F_{cd}\right)=0 \quad
{\rm and}\quad\nabla_b\left(\widehat X^{abcd}G_{cd}\right)=0\,.
\labell{eom2}
\end{equation}
For the action of interest \eqref{eqn4}, $X$ is given in
eq.~\eqref{eqn77} and at least in a regime where we treat $\gamma$ as
small, we can write
\begin{equation}
\left(X^{-1}\right)_{ab}{}^{cd}=I_{ab}{}^{cd} +8\gamma L^2 C_{ab}{}^{cd}
+O(\gamma^2)\,.
\labell{Xinverse}
\end{equation}
Further because of the traceless property of the Weyl tensor, one finds
\begin{equation}
\widehat X_{ab}{}^{cd}=\left(X^{-1}\right)_{ab}{}^{cd}
+O(\gamma^2)\,.
\labell{Xhat}
\end{equation}
With the change in sign of the order $\gamma$ contribution between
eqs.~\eqref{eqn77} and \eqref{Xinverse}, it is clear that our gauge
theory is no longer self-dual.

Actually given the planar black hole background \reef{eqn3}, it is
straightforward to calculate $X^{-1}$ exactly. First, we define a
six-dimensional space of (antisymmetric) index pairs with, \ie
$A,B\in\lbrace tx, ty, tu, xy, xu, yu\rbrace$ -- note both the ordering
of both the indices and the index pairs presented here. Then $X$ given
in eq.~\eqref{eqn77} becomes a diagonal six-by-six matrix
\begin{equation}
X_A{}^B={\rm diag}\left(1+\alpha, 1+\alpha, 1-2\alpha, 1-2\alpha,
1+\alpha, 1+\alpha \right)
\labell{Xdiag}
\end{equation}
where $\alpha=4 \gamma\,u^3$. Since $X$ is a diagonal matrix, $X^{-1}$
is also a diagonal matrix whose entries are simply the inverses of
those given in eq.~\eqref{Xdiag}. Note that $\alpha$ takes its maximum
value at the horizon $u=1$, \ie $\alpha_{max}=4\gamma$. Hence we must
constrain $-\frac14<\gamma<\frac18$ in order for the inverse to exist
everywhere in the region outside of the horizon. Of course, it is not a
coincidence that the effective Schr\"odinger equation in section
\ref{bond} became problematic (\ie the effective potentials contained a
pole) precisely outside of the same interval. In any event, the
physical regime \reef{eqn34} for $\gamma$ determined in section
\ref{bond} lies well within this range.

Using this notation and the background metric \reef{eqn3},
$\varepsilon_{ab}{}^{cd}$ becomes the following `anti-diagonal'
six-by-six matrix
\begin{equation}
\varepsilon_A{}^B=
\left[
\begin{matrix}
\ &\ &\ &\ &\ &\frac{r_0f}{L^2} \\
\ &\ &\ &\ &-\frac{r_0f}{L^2} &\ \\
\ &\ &\ &\frac{L^2}{r_0} &\ &\ \\
\ &\ &-\frac{r_0}{L^2} &\ &\ &\ \\
\ &\frac{L^2}{r_0f} &\ &\ &\ &\ \\
-\frac{L^2}{r_0f} &\ &\ &\ &\ &\ \\
\end{matrix}
\right]\,. \labell{antie}
\end{equation}
Combining these expressions, we can easily evaluate the duality
transformation \eqref{duality}, which is expressed using the new
notation as
\begin{equation}
F_{A}=g^2_4\,\left(X^{-1}\right)_{A}{}^B\,
\varepsilon_B{}^C\, G_C\,.
\labell{duality2}
\end{equation}
The final result is
\begin{eqnarray}
F_{tx}&=&\frac{g_4^2}{1+\alpha}\frac{r_0f}{L^2}G_{yu}\,,\quad
\ \ F_{ty}\ =\ -\frac{g_4^2}{1+\alpha}\frac{r_0f}{L^2}G_{xu}\,,
\nonumber\\
F_{tu}&=&\frac{g_4^2}{1-2\alpha}\frac{L^2}{r_0}G_{xy}\,,\quad\
F_{xy}\ =\ -\frac{g_4^2}{1-2\alpha}\frac{r_0}{L^2}G_{tu}\,,
\labell{duality3}\\
F_{xu}&=&\frac{g_4^2}{1+\alpha}\frac{L^2}{r_0f}G_{ty}\,,\quad
\ \ F_{yu}\ =\ -\frac{g_4^2}{1+\alpha}\frac{L^2}{r_0f}G_{tx}\,.
\nonumber
\end{eqnarray}
This duality transformation gives us a precise analytic relation
between the original gauge field $A_a$ and that, $B_a$, in the EM dual
theory. Of course, it would be less straightforward to express these
duality relations in a covariant construction using the Weyl curvature
tensor.

As discussed in \cite{sach}, from the perspective of the boundary field
theory, we can describe the CFT in terms of the original conserved
current $J_a$ (dual to the bulk vector $A_a$) or a new current
$\widehat{J}_a$ (dual to $B_a$). In the case of the Maxwell theory, the
EM self-duality means that both currents have identical correlators. In
the present case, where EM self-duality is lost, the correlators still
have a simple relation which is summarized by
 \bea
K^T(\omega,q)\ \widehat K^L(\omega,q)&=& 1 \,, \labell{slow1}\\
\widehat K^T(\omega,q)\  K^L(\omega,q)&=& 1 \,. \nonumber
 \eea
The detailed derivation for these relations can be found in appendix
\ref{appA}. The self-dual version of eq.~(\ref{slow1}), with $K =
\widehat K$, appeared in \cite{sach}. However, the conventions for the
EM duality transformation were different there, \ie they chose
$\hat{g}_4=g_4$. This choice changes the normalization of the dual
currents and so changes the constant on the right-hand-side of
eq.~\reef{slow1} to $(g_4)^{-4}$. In any event, these relations imply,
the longitudinal correlator in one theory is traded for the transverse
correlator in the dual theory, as reflected in eq.~\reef{duality3}.
Notably, eq.~(\ref{slow1}) has precisely the same form as that obtained
from general considerations of particle-vortex duality, but without
self-duality, in the condensed matter context, as we review in the
following subsection.

\subsection{Particle-Vortex Duality} \label{pavod}

Above, we discussed EM duality as a change of variables which allows us
to formulate the bulk theory in terms of two different gauge
potentials. This reformulation of the bulk theory implies that the
boundary CFT can be developed in terms of two `dual' sets of currents,
whose correlators are simply related using eq.~\reef{slow1}. As noted
in \cite{sach}, the latter is reminiscent of the structure of the
correlators in systems exhibiting particle-vortex duality. The
discussion there focused on self-dual examples, however, the latter is
an inessential feature to produce eq.~\reef{slow1}, as we illustrate
with the following simple example -- see also appendix B of
\cite{sach}.

Consider the field theory of a complex scalar $z$ coupled to a U(1)
gauge field
\begin{eqnarray}
\mathcal{S} &=& \int\!\! d^2\!x\, dt\; \Biggl[
\left|\left(\partial_\mu - i A_\mu\right) z \right|^2 + s |z|^2  + u
|z|^4  + \frac{1}{2e^2} \left(
\epsilon^{\mu\nu\lambda} \partial_\nu A_\lambda \right)^2   \Biggr] \ .
\label{sza}
\end{eqnarray}
We now look at the structure of the conserved U(1) currents of
$\mathcal{S}$, and their correlators. For simplicity, we will restrict
our discussion to $T=0$ to make the main point in the simplest context.
There is a natural generalization to $T>0$, which is needed to obtain
the full structure of the relationship in eq.~(\ref{slow1}), and which
was discussed in \cite{sach}.

The theory $\mathcal{S}$ has the obvious conserved U(1) current
\begin{equation}
J_\mu = \frac{1}{i} z^\ast \left[( \partial_\mu - i A_\mu) z \right]
-  \frac{1}{i} \left[ ( \partial_\mu - i A_\mu) z^\ast \right] z
\end{equation}
Because of current conservation, we can write
the two-point correlator of this current in the form (reminder, we are at $T=0$)
\begin{equation}
\left\langle J_\mu (p) J_\nu (-p) \right\rangle = \left(\delta_{\mu\nu}
- \frac{p_\mu p_\nu}{p^2} \right) \sqrt{p^2} K (p^2).
\label{e1}
\end{equation}
Here, we note that this correlator has been defined to be irreducible with respect to
the propagator of the photon, $A_\mu$.

The theory $\mathcal{S}$ has a second conserved U(1) current; this is the
`topological' current
\begin{equation}
\widehat J_\mu = \frac{1}{2 \pi}\, \varepsilon_{\mu}{}^{\nu\lambda}\, \partial_\nu A_\lambda
\label{hcurrent}
\end{equation}
We can interpret $\widehat J_\mu$ as the current of dual set of particles which are the
Abrikosov-Nielsen-Olesen vortices of the Abelian-Higgs model in eq~(\ref{sza}). Each such
vortex carries total $A_\mu$ flux of $2 \pi$, and hence the prefactor above. Indeed, there is
a dual formulation of the theory in eq (\ref{sza}) in which the vortices become the
fundamental complex scalar field $\widehat{z}$:
\begin{eqnarray}
\widehat{\mathcal{S}} &=& \int\!\! d^2\!x\, dt\; \Bigl[
\left|\partial_\mu \widehat{z} \right|^2 + \widehat{s} |\widehat{z}|^2  + \widehat{u}
|\widehat{z}|^4 \Bigr]
\label{hsza}
\end{eqnarray}
This dual theory has no gauge field because the vortices of
$\mathcal{S}$ only have short-range interactions. The particle number
current of this dual theory is the same as that in eq.~(\ref{hcurrent})
\begin{equation}
\widehat J_\mu = \frac{1}{i} \widehat{z}{\,}^\ast \partial_\mu \widehat{z}
-  \frac{1}{i}  \partial_\mu \widehat{z}{\,}^\ast \,  \widehat{z}.
\label{hcurrent2}
\end{equation}
Now, returning to the perspective of the original theory $\mathcal{S}$ in eq (\ref{sza}) and the
U(1) current in eq. (\ref{hcurrent}),
we can write the two-point correlator of $\widehat J_\mu$ in
the general form
\begin{equation}
\left\langle \widehat J_\mu (p)  \widehat J_\nu (-p) \right\rangle = \frac{1}{4 \pi^2}
\left(\delta_{\mu\nu} - \frac{p_\mu p_\nu}{p^2} \right) \frac{p^2}{p^2/e^2 - \Sigma (p^2)}
\label{e2a}
\end{equation}
where $\Sigma (p^2) $ is the photon self-energy.

The photon $A_\mu$ couples linearly to the current $J_\mu$, and so the photon self energy is
clearly the irreducible $J_\mu$ correlator, and so
\begin{equation}
\Sigma (p^2) = \sqrt{p^2} K (p^2)
\label{e2b}
\end{equation}
Also as $p^2 \rightarrow 0$ in IR, we have $\Sigma (p^2) \gg p^2 /e^2$
-- recall that here we are assuming the spacetime dimension $d=3$. So
we have
\begin{equation}
\left\langle \widehat J_\mu (p) \widehat J_\nu (-p) \right\rangle \simeq
- \left(\delta_{\mu\nu} - \frac{p_\mu p_\nu}{p^2} \right) \sqrt{p^2} \widehat K (p^2)
\label{e2}
\end{equation}
where from eqs.~(\ref{e2a}) and (\ref{e2b})
\begin{equation}
K (p^2) \widehat K(p^2) = \frac{1}{4 \pi^2}. \labell{slow2}
\end{equation}
This result is clearly the $T=0$ analog of eq.~(\ref{slow1}). It is
easily generalized to $T>0$, after separation into transverse and
longitudinal components, but we refrain from presenting those details
here.

\section{Discussion} \label{discuss}

Our main results for the frequency dependence of the conductivity
without self-duality were given in Fig.~\ref{fig1}, and we presented a
physical interpretation in Section~\ref{sec:intro}.
For $\gamma > 0$, the results had a qualitative similarity to
that expected from a Boltzmann transport theory of interacting particles,
while for $\gamma < 0$ the results resembled the Boltzmann transport
of vortices.

We will now discuss other aspects of these results.
We also see from Fig.~\ref{fig1} that the large frequency limit is
unaffected by the new coupling, \ie $\sigma(\omega=\infty) =1/g^2_4$.
We can understand this result from the fact that the Weyl curvature
vanishes in the asymptotic region of the black hole region and so the
new interaction in eq.~\reef{eqn4} has no effect there.

Further, we have
 \be
\frac{\sigma(\omega=0)}{\sigma(\omega=\infty)}=1+4\gamma
 \labell{result0}
 \ee
and so this ratio varies between 4/3 and 2/3 in the allowed physical
regime given in eq.~\reef{eqn34}. Thus the allowed range of variation
in the conductivity by non-self-duality is smaller than 33\% and can
have either sign, in our model. This should be contrasted from the
large variation obtained from the weak-coupling Boltzmann analyses. In
the $\epsilon=4-d$ expansion (where $d$ is the spacetime dimension), it
was found that generically \cite{subir1}
  \be \frac{\sigma(\omega=0)}{\sigma
(\omega=\infty)} \sim \frac{1}{\epsilon^2}\,.
 \ee
Similarly, in the large $N$ expansion (where $N$ is the number of
components of a vector (and not matrix) field), we have \cite{subir2}
 \be
\frac{\sigma(\omega=0)}{\sigma (\omega=\infty)} \sim N\,.
 \ee
In both cases, the ratio becomes large in the regime of applicability
of the analysis. Thus the AdS/CFT analysis gives a useful result for
this ratio in the complementary limit of very strong interactions.

Also note that the conductivity in Fig.~\ref{fig1} does not vary
monotonically, rather it seems there is an extremum at $\omega\simeq
2\pi T$. For $\gamma >0$, this oscillation is as anticipated from
Drude-like considerations of particle transport in ref.~\cite{subir1},
and for $\gamma < 0$ we argued in Section~\ref{sec:intro} that such an
oscillation is obtained from Drude-like vortex transport.
Recall that in the AdS/CFT correspondence, particle-vortex duality in
the boundary theory is realized as EM duality in the bulk theory. Hence
we can make the previous point explicit for our holographic model using
the formalism developed in Section \ref{duality1}. That is, for any
given value of $\gamma$, we can explicitly construct the EM dual theory
and evaluate the conductivity. In Fig.~\ref{fig99x}, we have plotted
the resulting conductivities for the original bulk theory and the EM
dual theory for $\gamma=\pm 1/12$. As expected, for $\gamma=-1/12$, the
conductivity of the dual theory exhibits a Drude-like peak at small
$\omega$. For $\gamma=1/12$, a similar peak appears for the original
theory while the EM dual theory exhibits a dip in the conductivity at
small $\omega$. For either value of $\gamma$, the figure also
illustrates that the conductivities of the two dual theories are not
precise inverses of one another, except for $\omega\to 0,\,\infty$.
This occurs because the function $K^T(\omega,q=0)$ is only precisely
real in the latter limits.
\FIGURE{
\includegraphics[width=11cm,height=7cm]{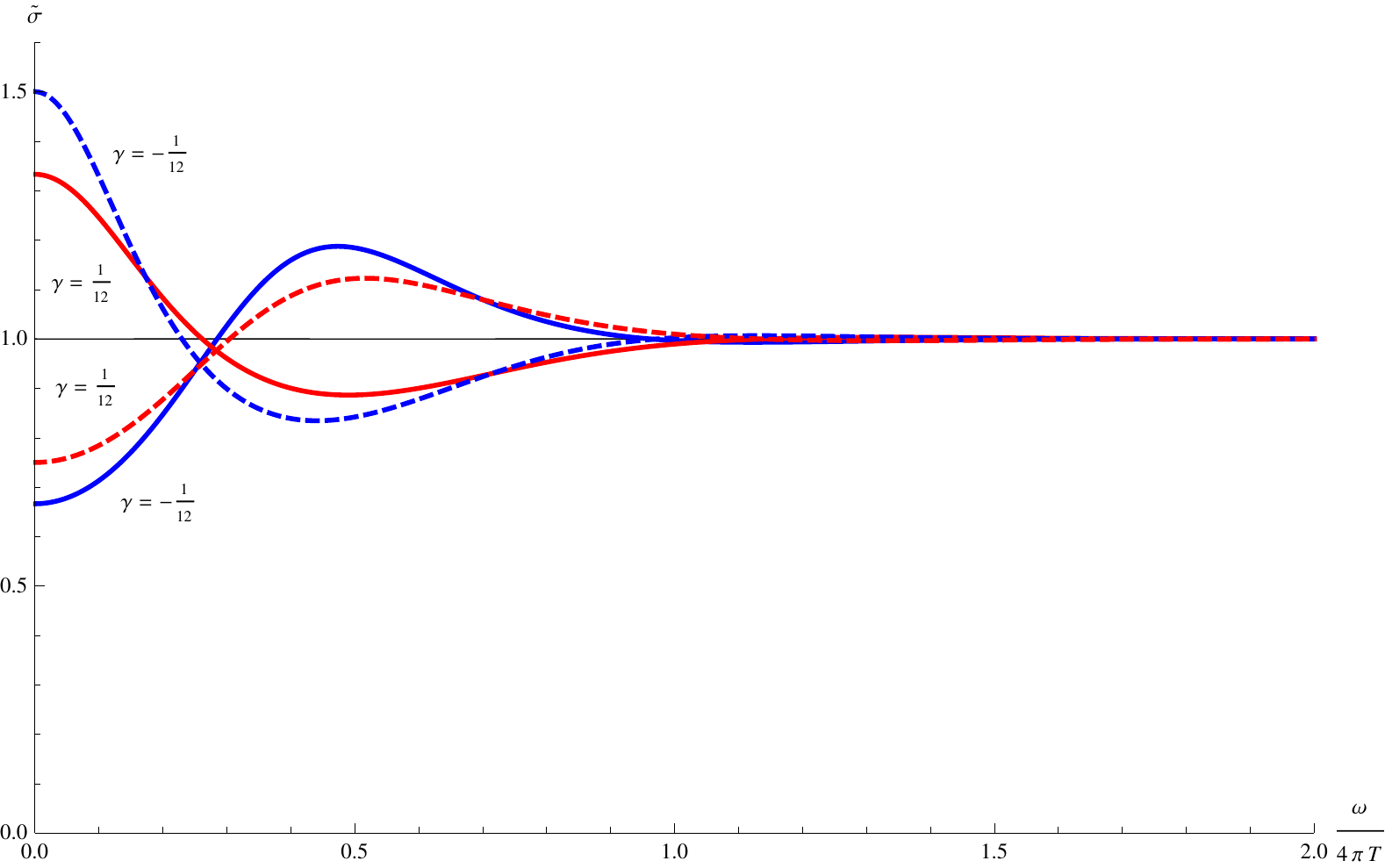}
\caption{The (dimensionless) conductivity $\tilde\sigma=g_4^2\sigma$ is
plotted versus the (dimensionless) frequency $\wn=\omega/(4\pi T)$ for
various values of $\gamma$. The solid curves correspond to the same
conductivities displayed previously in Fig.~\ref{fig1} --- red for
$\gamma=1/12$ and blue for $\gamma=-1/12$. The dashed curves show the
conductivity calculated from the EM dual theory for the same values of
$\gamma$.} \label{fig99x}}

Oscillations in the conductivity similar to those in Fig.~\ref{fig1}
were observed in \cite{wapler}. The latter studied the transport
properties of currents on a three-dimensional defect immersed in the
thermal path of a four-dimensional superconformal gauge theory. The
holographic bulk theory consisted of probe D-branes embedded in
AdS$_5\times S^5$ and the oscillations were an effect of stringy
corrections to the usual D-brane action. Implicitly, the
four-derivative interaction considered there would have been a linear
combination of the $\alpha_{5,6,7}$ terms in eq.~\reef{act14}. In this
previous setting, the calculations were perturbative and the
oscillatory contribution to the conductivity was suppressed by a factor
of $\lambda^{-1/2}$ relative to the constant term produced by the
Maxwell action on the brane -- as usual, $\lambda$ denotes the `t Hooft
coupling of the four-dimensional gauge theory.

Section \ref{pree} provided some motivation for introducing the new
four-derivative interaction in eq.~\reef{eqn4}. However, there was a
certain liberty in choosing the precise form of the curvature in this
interaction. From a certain perspective, the following vector action
may be preferred:
\begin{equation}
I'_{vec}=\frac{1}{\tg^2_4} \int d^4x \sqrt{-g}\left[-\frac{1}{4}F_{ab}F^{ab} +
\alpha L^2 \left[ R_{abcd}F^{ab}F^{cd}-4 R_{ab}F^{a c}F^b{}_{c}
 +R F^{ab}F_{ab} \right] \right]\,.
\labell{eqn35}
\end{equation}
The advantage of the higher-derivative term above is that it produces
second-order equations of motion for both the gauge field and metric in
any general background. We can think of this term arising from
Kaluza-Klein reduction of Gauss-Bonnet gravity in five-dimensional
space-time \cite{nonpub}. Now the generalized Maxwell's equations are
\begin{equation}
\nabla_{a} \left[ F^{ab} -4 \alpha L^2(R^{ab}{}_{cd}F^{cd} - 2R^{ac}F_{c}{}^{b} +
2 R^{bc} F_c{}^{ a} +R F^{ab})\right] = 0\,.
\labell{eqn36}
\end{equation}

Before considering the charge transport for this theory, we note that
AdS vacuum and the neutral black hole \reef{eqn1} remain unmodified
with this choice of the four-derivative interaction. In particular
then, for the black-hole background, we still satisfy the vacuum
Einstein equations, \ie $R_{ab}=-3/L^2\,g_{ab}$. Further, the Reimann
curvature tensor $R_{abcd}$ is related to the Weyl tensor $C_{abcd}$ by
\begin{equation}
R_{abcd}=C_{abcd}+g_{a[c}R_{d]b}-g_{b[c}R_{d]a}-\frac{1}{3}R\,g_{a[c}g_{d]b}\,.
 \labell{eqn38}
\end{equation}
By substituting these relations into eq.~\eqref{eqn35}, we find that
the action becomes
\begin{equation}
I'_{vec}=\frac{1+8\alpha}{\tg^2_4} \int d^4x \sqrt{-g}\left(-\frac{1}{4}F_{ab}F^{ab} +
 \frac{\alpha}{1+8\alpha} L^2 C_{abcd}F^{ab}F^{cd} \right)\,.
\labell{eqn39}
\end{equation}
Hence, this expression for action is identical to eq.~\eqref{eqn4} if
we identify the couplings:
 \be
 g_4^2=\frac{\tg^2_4}{1+8\alpha}\quad{\rm and}\quad
 \gamma=\frac{\alpha}{1+8\alpha}
 \labell{rift}
 \ee
Hence in the neutral plasma, all of the charge transport properties of
the new theory are identical to those found in the main text, as long
as we make this identification of the couplings in the bulk gauge
theory. For example, we have explicitly applied the analysis of section
\ref{bond} to the new action \reef{eqn35} and found this produces the
constraints $-1/20\leq\alpha \leq 1/4$. One can easily verify that this
range precisely matches that in eq.~\reef{eqn34} for $\gamma$ using the
identification of the gauge theory couplings in eq.~\reef{rift}.

It would be interesting to examine charged black holes in this new
theory \reef{eqn35}. Beyond analyzing the effects of adding a chemical
potential in the boundary CFT, it would be interesting to examine the
so-called ``entropy problem" in this theory. That is, at zero
temperature, charged black holes still have a finite horizon area for
the Einstein-Maxwell theory in the bulk and hence the dual CFT has a
large entropy even at $T=0$ but nonvanishing chemical potential. It
would be interesting to determine how this feature found in simple
holographic CFT's is affected by the introduction of the new higher
derivative bulk interaction in eqs.~\reef{eqn4} or \reef{eqn35}. Such
investigations would require numerical work that would be greatly
facilitated by having second-order equations, as produced by the above
action \reef{eqn35}.

As discussed in the introduction, we are following a program of
expanding the universality class of the holographic CFT by introducing
new higher-derivative interactions to the bulk action. The simplest way
to characterize the effect of the new interactions is to examine the
changes which are produced in the vacuum $n$-point functions in the
CFT. As alluded to above, the Weyl curvature vanishes in AdS space and
so we may infer that in the vacuum of the dual CFT (with vanishing
temperature and charge density), there are no changes to any of
two-point functions, \ie $\langle J_a(x) J_b(y)\rangle_0$ and $\langle
T_{ab}(x) T_{cd}(y)\rangle_0$, where the subscript 0 indicates the
two-point functions are evaluated in the vacuum or at $T=0$. That is,
the two-point functions are independent of $\gamma$ in the vacuum. In
particular then, the charge transport properties of the holographic CFT
must be independent of $\gamma$ at $T=0$. On the other hand, recall the
simple $\gamma$ dependence which appears in eq.~\reef{eqn5x3} for the
conductivity at $\omega=0$. Clearly, this means that the limits,
$T\to0$ and $\omega\to0$, do not commute, as was also emphasized in
ref.~\cite{subir1}.

As described in \cite{Hofman1,Hofman2}, the key effect of the new bulk
interaction \reef{eqn4} is to modify the three-point correlator
$\langle T_{ab}(x) J_c(y) J_d(z)\rangle_0$. One can show that in any
CFT, conformal symmetry will completely fix this three-point function
between the stress tensor and two conserved currents up to two constant
parameters \cite{ozzy}. One of these parameters vanishes in the
holographic dual of an Einstein-Maxwell theory. However, this extra
parameter is nonvanishing for the CFT dual for our extended theory with
$\gamma\ne0$.
In particular, as discussed in section \ref{bond}, the parameter $a_2$
in eq.~\reef{eqn34x1} is only nonvanishing in the boundary CFT when
$\gamma\ne0$.
Of course in a thermal bath, the expectation value of the stress tensor
is nonvanishing. Hence it should be possible to use the previous
three-point function to infer the leading $\gamma$ modification to the
two-point correlator $\langle J_a J_b \rangle_T$ at finite temperature,
\eg with an approach similar to that considered in \cite{truck}. In
principle then, such a (perturbative) calculation in the CFT should
already indicate that self-duality is lost.

Above, we discussed the behavior of the conductivity, which is related
to the current correlator at zero momentum. We also studied the full
momentum dependence of these correlators and obtained the duality
relation in eq.~(\ref{slow1}), which applied in the general case
without {\em self\/}-duality. Remarkably, this has the same form as
that obtained by applying particle-vortex duality to a
(2+1)-dimensional field theory of a single complex scalar, as we
reviewed in section~\ref{pavod}: note that this theory is not self-dual
(and self-duality is not expected in general, except for a particular
theory with two complex scalar fields \cite{sach,vish}). In the single
scalar field case, as discussed in section \ref{pavod}, $K^{T,L}$
characterize the transverse/longitudinal components of the two-point
correlations of the current of the scalar particles -- see
eq.~\reef{green1} -- while $\widehat{K}^{T,L}$ characterize the
corresponding quantities of the vortex current.

Of course, the constants on the right-hand-side of eqs.~\reef{slow1}
and \reef{slow2} are seen to be different. In both cases, this constant
depends on the conventions used to normalize the currents and a new
normalization would change the constant in either model. Hence one may
ask if these relations can be expressed in a way which removes this
ambiguity. As we will show, one possibility is to replace
eq.~\reef{slow1} by
 \bea
K^T(\omega,q)\ \widehat K^L(\omega,q)&=& \sigma_0\,\hat{\sigma}_0 \,,
 \labell{slow1a}\\
\widehat K^T(\omega,q)\  K^L(\omega,q)&=& \sigma_0\,\hat{\sigma}_0 \,,
 \nonumber
 \eea
where $\sigma_0$ is the conductivity at zero momentum and zero
frequency and $\hat{\sigma}_0$ is the same quantity for the dual
currents. For our holographic model, $\sigma_0$ was given in
eq.~\reef{eqn5x3} and given the discussion in section \ref{duality1},
it is a simple exercise to show that $\sigma_0\,\hat{\sigma}_0=1$.
Hence, in this case, we easily recover eq.~\reef{slow1} from
eq.~\reef{slow1a} above. However, the latter equation applies quite
generally as we will now show: First, given the expressions for
$\sigma$ and $K^T$ in eqs.~\reef{eqn16} and \reef{apeq6}, respectively,
it is straightforward to show that $\lim_{\omega\to0}\, K^T(\omega,q=0)
=\sigma_0$. Further with vanishing momentum,
$K^L(\omega,q=0)=K^T(\omega,q=0)$ and hence we also have
$\lim_{\omega\to0}\, \widehat{K}^T(\omega,q=0) =\hat{\sigma}_0\,.$ Now
if we know that the product $K^T\, \widehat{K}^L$ is constant, we can
evaluate the constant at vanishing momentum and vanishing frequency and
then our discussion leads us to write eq.~\reef{slow1a}. This
expression will apply independent of the conventions used to normalize
the currents and applies equally well for the field theory examples
considered in section \ref{pavod} and in \cite{sach} as for our
holographic model.

The holographic relation of EM duality in the bulk and particle-vortex
duality in the boundary theory was first noted in \cite{sach,wits} and
the effect of this bulk transformation on the boundary transport
properties was further studied in \cite{sach2} -- see also
\cite{wapler,perk}. Particle-vortex duality can be extended to an
$SL(2,Z)$ action on three-dimensional CFT's \cite{wits,cbbd} and the
holographic realization of these group transformations on the bulk
theory was discussed in \cite{wits}. In particular, the $S$
transformation corresponds to applying EM duality in the bulk. To
discuss the $T$ transformation, the bulk action must be extended to
include a $\theta$-term and acting with the $T$ generator corresponds
to making a $2\pi$ shift of $\theta$. Of course, implicitly or
explicitly, the previous holographic discussions assumed a standard
Maxwell action for the bulk vector. It would be interesting to extend
this discussion of the full $SL(2,Z)$ action to the generalized action
\reef{general} introduced in section \ref{diff}. Associating the $S$
generator with EM duality as in \cite{wits}, one can easily verify that
$S^2=-1$ using eq.~\reef{duality}. To include the $T$ generator, we
would need generalize $X$ to include parity violating terms, \ie
nonvanishing $z_2(u)$ and $z_3(u)$ in eq.~\reef{Xround}. We leave this
as an interesting open question.

To close, we wish to emphasize that our investigation here has
considered a simple toy model and one should be circumspect in
interpreting the results of our analysis. While string theory will
generate the higher derivative interactions in our action \reef{eqn4},
it certainly also produces many other higher order terms which
schematically take the form $R^n\,F^2$. For example, some such terms,
were explicitly constructed (amongst many others) and studied in
\cite{rnf2}. Any terms with this schematic form would still fall in the
class of our general action \reef{general} and so modify the charge
transport properties in a similar way. A key feature of our model was
that we were able to identify physical restrictions which constrained
the new coupling $\gamma$ to fall in relatively narrow range
\reef{eqn34}. As a result, the conductivity remained relatively close
to the self-dual value. Our expectation is that similar restrictions
appear for general string models, however, finding more comprehensive
physical constraints in this context remains an interesting open
question \cite{open}. As seen here and elsewhere
\cite{Hofman1,Buchel0,shank,Hofman2}, the interplay between the
boundary and bulk theories in the AdS/CFT correspondence is beginning
to provide new insights into this question.

\vskip 2ex
 \noindent {\bf Acknowledgments:} We thank Cliff Burgess, Jaume Gomis,
Sean Hartnoll, Brandon Robinson, Brian Sheih and Aninda Sinha for
discussions.
SS would like to thank T.~Senthil for noting the connection
between vortex transport and the $\gamma < 0$ conductivity.
AS would like to thank Jorge Escobedo, Yu-Xiang Gu and
Sayeh Rajabi for fruitful and stimulating discussions. Research at
Perimeter Institute is supported by the Government of Canada through
Industry Canada and by the Province of Ontario through the Ministry of
Research \& Innovation. RCM also acknowledges support from an NSERC
Discovery grant and funding from the Canadian Institute for Advanced
Research. SS is grateful to the Perimeter Institute for hospitality,
and acknowledges support by the National Science Foundation under grant
DMR-0757145, by the FQXi foundation, and by a MURI grant from AFOSR.

\appendix

\section{Retarded Green's functions and EM duality} \label{appA}

In this appendix, we find the retarded Green's functions of currents in
the boundary field theory for finite frequency and finite momentum and
further we examine the relationship between the Green's functions in
the two theories related by EM duality in the bulk. In this discussion,
we work with the general vector action \reef{general} and its EM dual
\reef{eqn67xx}. Recall that the relation between coefficients $X$ and
$\widehat X$ appearing in these two actions is given in
eq.~\reef{kinetic} and the field strengths in the two theories are
given by $F_{ab}\equiv\partial_a A_b-\partial_b A_a$ and
$G_{ab}\equiv\partial_aB_b-\partial_bB_a$, respectively. Further the
duality relation between these two field strengths is given in
eq.~\reef{duality}.

For simplicity, we will begin by assuming that $X_{ab}{}^{cd}$ is
diagonal in the six-dimensional space defined by the antisymmteric
index pairs
\be
A,B\in\lbrace tx, ty, tu, xy, xu, yu\rbrace\,.
\labell{indices}
\ee
This property holds for the specific theory \reef{eqn4} studied in the
main text, as shown in eq.~\reef{Xdiag}. We comment on more general
cases at the end of the appendix. Given this assumption, we write
\begin{equation}
X_A{}^B={\rm diag}\left( X_1(u),X_2(u),X_3(u),X_4(u),X_5(u),X_6(u)
\right)\,.
\labell{Xdiag1}
\end{equation}
Further rotational symmetry in the $xy$-plane would restrict this
ansatz with $X_1(u)=X_2(u)$ and $X_5(u)=X_6(u)$. However, we leave this
symmetry as implicit, since it is not required in the following. Now
the inverse\footnote{We assume that the functions $X_i$ remain finite
and positive throughout $u\in [0,1]$ in order that $X_A{}^B$ is
invertible and the bulk propagators for the gauge potential are
well-behaved there.} $X^{-1}$ is simply the diagonal matrix with
entries $1/X_i(u)$ and, given eq.~\eqref{kinetic}, $\widehat X_A{}^B$
is also diagonal with
\bea
\widehat{X}_A{}^B&=&{\rm diag}\left(
\widehat{X}_1(u),\widehat{X}_2(u),\widehat{X}_3(u),
\widehat{X}_4(u),\widehat{X}_5(u),\widehat{X}_6(u))\right)
\nonumber\\
&=&{\rm diag}\left(
\frac{1}{X_6(u)},\frac{1}{X_5(u)},\frac{1}{X_4(u)},\frac{1}{X_3(u)},
\frac{1}{X_2(u)},\frac{1}{X_1(u)}\right)\,.
\labell{hXdiag}
\eea

Now we review the general structure of the Green's functions in the
boundary theory, from the discussion in \cite{sach}. Together current
conservation and spatial rotational invariance -- Lorentz invariance is
lost with $T\ne0$ -- dictate the form of the retarded Green's functions
as
\be
\mathcal{G}_{\mu\nu}({\bf q})=\sqrt{{\bf q}^2}\left( P^T_{\mu\nu}\,
K^T(\omega,q)+ P^L_{\mu\nu}\, K^L(\omega,q) \right) \,.
\labell{green1}
\ee
where we use the notation: ${\bf q}^\mu=(\omega,q^x,q^y)$,
$q^2=[(q^x)^2+(q^y)^2]^{1/2}$ and ${\bf q}^2=q^2-\omega^2$. Further,
$P^T_{\mu\nu}$ and $P^L_{\mu\nu}$ are orthogonal projection operators
defined by
\be
P^T_{tt}=0=P^T_{ti}=P^T_{it} \,,\quad P^T_{ij}=\delta_{ij}-\frac{q_i
q_j}{q^2} \,,\quad P^L_{\mu\nu}=\left(\eta_{\mu\nu}-
\frac{q_{\mu}q_{\nu}}{|\bf{q}|^2}\right) - P^T_{\mu\nu}\,,
\labell{apeq5}
\ee
with $i,\,j$ denoting spatial indices while $\mu\,,\nu$ run over both
space and time. If, for simplicity, we choose ${\bf
q}^\mu=(\omega,q,0)$, then we have
\be
\mathcal{G}_{yy}(\omega,q)=\sqrt{q^2 - \omega^2}\,K^T(\omega,q) \, ,
\qquad \mathcal{G}_{tt}(\omega,q)=-\frac{q^2}{\sqrt{q^2-\omega^2}}\,K^L
(\omega,q)\,.
\labell{apeq6}
\ee
Of course, this general structure applies for both boundary theories,
that is, both for the theory dual to the vector potential $A_a$ and
that dual to $B_a$. Our notation will be that the above expressions
refer to the theory dual to $A_a$ while
$\widehat{\mathcal{G}}_{\mu\nu}$, $\widehat K^T$ and $\widehat K^L$ are
the corresponding expressions for the boundary currents dual to $B_a$.

The first step in the holographic calculation of the Green's functions
is to solve the bulk equations of motion. Hence we begin as in section
\ref{cond} by taking a plane-wave ansatz \eqref{eqn6} for $A_a$ and
$B_a$. Further, we choose ${\bf q}^\mu=(\omega,q,0)$ and work in radial
gauge with $A_u(u,{\bf q})=0=B_u(u,{\bf q})$. With these choices and
the background metric \reef{eqn3}, the $A_a$ equations of motion
become:
\begin{align}
 A_t' + \frac{q\,f}{\omega}\frac{X_5}{X_3} A_x' & \;=\;0 \labell{apeq4} \\
A_t''+\frac{X_3'}{X_3} A_t' -  \frac{L^4}{r_0^2}\frac{q}{f}\frac{X_1}{X_3}
 (q A_t+\omega  A_x) & \;=\;0 \labell{apeq1}\\
A_x''+\left(\frac{X_5'}{X_5}+\frac{f'}{f}\right)A_x'
+\frac{L^4}{r_0^2}\frac{\omega}{f^2}
\frac{X_1}{X_5} (q A_t+\omega  A_x) & \; = \;0 \labell{apeq2}\\
A_y''+\left(\frac{X_6'}{X_6}+\frac{f'}{f}\right)A_y'-\frac{L^4}{r_0^2}
\frac{\omega^2 X_2-q^2 f
X_4}{f^2 X_6} A_y & \;=\;0 \labell{apeq3}
\end{align}
where we recall that $f=1-u^3$. For the EM dual gauge theory, the
equations of motion are given by simply replacing $A_a\to B_a$ and
$X_i\to\widehat{X}_i$ in the expressions above.

In general, there are two independent physical modes for the
four-dimensional bulk gauge field. Above, we see that $A_y$ decouples
in eq.~\reef{apeq3} to provide one of these modes, while $A_t$ and
$A_x$ are coupled in the remaining equations. Of course, the analogous
results apply to $B_a$ in the EM dual theory. Now explicitly writing
out the duality relations \reef{duality2} in the present case, we find
\begin{eqnarray}
F_{tx}&=&\frac{g_4^2}{X_1}\frac{r_0f}{L^2}\,G_{yu}\,,\quad
F_{ty}\ =\ -\frac{g_4^2}{X_2}\frac{r_0f}{L^2}\,G_{xu}\,,
\nonumber\\
\quad
F_{tu}&=&\frac{g_4^2}{X_3}\frac{L^2}{r_0}\,G_{xy}\,,\quad\
F_{xy}\ =\ -\frac{g_4^2}{X_4}\frac{r_0}{L^2}\,G_{tu}\,,
\labell{duality4}\\
F_{xu} &=& \frac{g_4^2}{X_5}\frac{L^2}{r_0f}\,G_{ty}\,,\quad
F_{yu}\ =\ -\frac{g_4^2}{X_6}\frac{L^2}{r_0f}\,G_{tx}\,. \nonumber
\end{eqnarray}
Hence, at a schematic level, EM duality exchanges the $A_y$ mode for
that in $B_{t,x}$ and similarly the $A_{t,x}$ and $B_y$ are exchanged.
Given the holographic relationship between the bulk and boundary
theories, we expect that there are connections between the Green's
functions, $\mathcal{G}_{\mu\nu}$ and $\widehat{\mathcal{G}}_{\mu\nu}$,
generalizing those found in \cite{sach}. However, given the previous
observation, more specifically, $\mathcal{G}_{yy}$ should be related to
$\widehat{\mathcal{G}}_{tt}$ (as well as $\widehat{\mathcal{G}}_{xx}$
and $\widehat{\mathcal{G}}_{tx}$) and similarly $\mathcal{G}_{tt}$, to
$\widehat{\mathcal{G}}_{yy}$.

To develop these connections in detail, we must extend the holographic
calculation of the Green's functions given in section \ref{cond} to
include the mixing between $A_t$ and $A_x$, noted above. First, we
solve the equations of motion \eqref{apeq4}-\eqref{apeq3} for $A_\mu$
with infalling boundary conditions at the horizon and asymptotic
boundary conditions: $\lim \limits_{u \to 0} A_{\mu}=A_\mu^0$. To
account for mixing between different components of the gauge potential,
we may write \cite{sach}: $A_\mu(u)\; =\; M_{\mu}{}^{\nu}(u)
A^0_{\nu}$. Now, substituting the solutions into the action
\eqref{general} and integrating by parts leaves an surface term at the
asymptotic boundary, which generalizes that given in eq.~\reef{eqn18},
 \be
I_1\;=\;\frac{2\pi T}{3g_4^2} \int d^3x\, \left[ X_3\, A_t A_t'-X_5\,
A_x A_x' - X_6\, A_y A_y' \right]_{u\to0}\,.
 \labell{apeq7}
 \ee
After Fourier transforming in the boundary directions, we extract the
desired Green's functions as
 \bea
\mathcal{G}_{tt}(\omega,q)&=&\frac{4\pi T}{3g_4^2}X_3(0) \left.
\frac{\delta A'_t(u)}{\delta A_t^0}\right|_{u\to0}\,, \labell{ggtt}\\
\mathcal{G}_{xx}(\omega,q)&=&-\frac{4\pi T}{3g_4^2}X_5(0) \left.
\frac{\delta A'_x(u)}{\delta A_x^0}\right|_{u\to0}\,, \labell{ggxx}\\
\mathcal{G}_{tx}(\omega,q)&=&\frac{2\pi T}{3g_4^2}\left[X_3(0)
\frac{\delta A'_t(u)}{\delta A_x^0} -X_5(0)
\frac{\delta A'_x(u)}{\delta A_t^0}\right]_{u\to0}\,,\labell{ggtx}\\
\mathcal{G}_{yy}(\omega,q)&=&-\frac{4\pi T}{3g_4^2}X_6(0) \left.
\frac{\delta A'_y(u)}{\delta A_y^0}\right|_{u\to0} \,.
 \labell{ggyy}
 \eea
Here we have used that the equations of motion
(\ref{apeq4}--\ref{apeq3}) only mix $A_t$ and $A_x$. One may also
easily verify that eq.~\reef{ggyy} reduces to the expression in
eq.~\reef{eqn21} when $X_6(0)=1$, as in the main text.

Next consider the Green's functions $\mathcal{G}_{yy}$. Assume that we
have $A_y(u)=\psi(u)A_y^0$ where $\psi(u)$ is a solution of
eq.~\eqref{apeq3} satisfying the appropriate boundary conditions. In
particularly, the asymptotic normalization is $\psi(u=0)=1$. Then from
\eqref{ggyy}, we have
 \be
\mathcal{G}_{yy}(\omega,q) = -\frac{4 \pi T}{3g_4^2}
\,X_6(0)\,\psi'(0)\,.
 \labell{gyy}
 \ee
Given that EM duality exchanges $A_y$ with $B_{t,x}$, we now look for a
relation between this result and that for $\widehat{\mathcal{G}}_{tt}$.
From the expression for $F_{xy}$ in eq.~\reef{duality4}, we find
$B_t'(u)\propto X_4(u)\,A_y(u)$ and so $X_4(u)\,\psi(u)$ provides a
solution of the equations of motion for $B_t'(u)$ in the EM dual
theory. While it is clear that the required infalling boundary
condition is satisfied at the horizon with $\psi(u)$,  we must expect
that the normalization has to be adjusted in order to satisfy the
desired asymptotic boundary condition. Hence we introduce a new
constant $C_1$ setting $B_t'(u)=C_1\, X_4(u)\,\psi(u)$. In order to fix
this constant, we consider the analog of eq.~\reef{apeq1} in the EM
dual theory and take the limit $u\to0$ to find
 \be
 C_1=\frac{L^4}{r_0^2} \frac{q(q\,B_t^0+\omega\,B_x^0)}{X_6(0)\,\psi'(0)}
 \labell{apeq10}
 \ee
where deriving this expression uses $\widehat{X}_1=1/X_6$ and
$\widehat{X}_3=1/X_4$. Now the EM dual counterpart of eq.~\eqref{ggtt}
yields

 \be
\widehat{\mathcal{G}}_{tt}(\omega,q)= \frac{4\pi
T}{3\hat{g}_4^2}\widehat{X}_3(0) \left. \frac{\delta B'_t(u)}{\delta
B_t^0}\right|_{u\to0}=\frac{3{g}_4^2}{4\pi T}\,
\frac{q^2}{X_6(0)\,\psi'(0)}\,. \labell{emggtt}
 \ee
Here we have used the relations: $\hat{g}_4=1/g_4$ and $r_0/L^2=4\pi
T/3$. Hence, combining eqs.~\eqref{gyy} and \reef{emggtt}, we find
 \be
\mathcal{G}_{yy}(\omega,q)\, \widehat{\mathcal{G}}_{tt}(\omega,q)=- q^2
\,.
 \labell{apeq13}
 \ee
Further, using eq.~\eqref{apeq6}, this relation can be written as
 \be
K^T(\omega,q) \, \widehat{K}^L(\omega,q)=1 \,.
 \labell{apeq14}
 \ee

Now it is clear that the EM dual version of the above discussion would
follow through without change. That is, we would begin by constructing
an expression for $\widehat{\G}_{yy}$ analogous to eq.~\reef{gyy} and
then the counterpart of eq.~\reef{emggtt} for ${\G}_{tt}$. The final
result emerging from these results would then be
\be
\widehat{K}^T(\omega,q) \, K^L(\omega,q)=1 \,.
\labell{apeq17}
\ee

To close our discussion, we comment on more general cases where $X$
contains off-diagonal terms. To begin, let us write the most general
tensor which is consistent with rotational symmetry in the $xy$-plane:
\begin{equation}
X_A{}^B=
\left[
\begin{matrix}
X_1(u) &\ &\ &\ &\frac{r_0f}{L^2}z_1(u) &\frac{r_0f}{L^2}z_2(u) \\
\ &X_1(u) &\ &\ &-\frac{r_0f}{L^2}z_2(u) &\frac{r_0f}{L^2}z_1(u) \\
\ &\ &X_3(u) &\frac{L^2}{r_0}z_3(u) &\ &\ \\
\ &\ &-\frac{r_0}{L^2}z_3(u) &X_4(u) &\ &\ \\
-\frac{L^2}{r_0f}z_1(u) &\frac{L^2}{r_0f}z_2(u) &\ &\ &X_5(u) &\ \\
-\frac{L^2}{r_0f}z_2(u) &-\frac{L^2}{r_0f}z_1(u) &\ &\ &\ &X_5(u) \\
\end{matrix}
\right]\,, \labell{Xround}
\end{equation}
where we are using the notation introduced in eq.~\reef{indices}, as
well as the background metric \reef{eqn3}. Note the pre-factors in the
off-diagonal terms reflect the tensor structure of $X_{ab}{}^{cd}$,
which is slightly obscure in this notation, \eg $X_5{}^1 =
g_{xx}\,g_{uu}\, g^{tt} g^{xx}\,X_1{}^5= -L^4/(r_0f)^2\,X_1{}^5$. Now,
as noted above, rotational invariance imposes two relations on the
diagonal entries, \ie $X_2=X_1$ and $X_6=X_5$. However, as shown above,
this symmetry is remarkably restrictive on the off-diagonal components
as well and our general tensor \reef{Xround} only contains three
independent terms amongst all of the possible entries. Now, if we
further demand that this background tensor preserves parity, we must in
fact set $z_2(u)=0=z_3(u)$ and we are left with only one function
$z_1(u)$ determining all of the allowed off-diagonal components. Note
that these remaining off-diagonal terms preserve parity but violate
time-reversal invariance.\footnote{The $z_2(u)$ and $z_3(u)$ terms
violate both parity and time-reversal invariance.}

If we restrict ourselves to the parity invariant case, it is
straightforward to generalize our previous discussion to accommodate
the general $X$ above (with $z_2=0=z_3$). Although the intermediate
expressions are somewhat more involved, we find that the final Green's
functions still satisfy eqs.~\reef{apeq14} and \reef{apeq17}.

Note that parity invariance was implicit in the decomposition of the
Green's functions in eq.~\reef{green1}. If parity violating terms were
allowed there would be an additional contribution of the form.
\be
\Delta\G_{\mu\nu}=i\,\varepsilon_{\mu\nu\sigma}q^\sigma\,
K^P(\omega,q)\,.
\labell{green2}
\ee
Hence the present analysis must be revised to accommodate these parity
violating terms. Our expectation is that particle-vortex duality still
provides relations between the three functions $K^T$, $K^L$ and $K^P$,
describing the Green's functions of the two dual theories.
A preliminary examination of the equations of motion and the EM duality
relations suggests that, in this general case, $K^T$, $K^L$, $K^P$ and
their dual counterparts should satisfy three relations. However, the
details of this interesting case are left as an open problem for future
work.

\end{document}